\documentclass[aps,prb,reprint,superscriptaddress,twocolumn]{revtex4-2}

\usepackage{bm,mathrsfs,dcolumn,graphicx,color} 
\usepackage{amsmath}
\usepackage{graphicx}
\usepackage[T1]{fontenc}
\usepackage{hyphenat}
\usepackage{setspace}
\usepackage{url}
\usepackage{xcolor}
\usepackage[utf8]{inputenc}

\usepackage{amssymb}
\usepackage{epsfig}
\usepackage[english]{babel}
\usepackage{empheq}
\usepackage[bbgreekl]{mathbbol}

\definecolor{myblue}{rgb}{.8, .8, 1} 

\def\dfrac{\displaystyle\frac}

\newcommand{\bra}[1]{\left< #1 \right|}
\newcommand{\ket}[1]{\left| #1 \right>}

\newcommand{\br}{\mathbf{r}}
\newcommand{\bk}{\mathbf{k}}

\newcommand{\bq}{\mathbf{q}}

\newcommand{\bp}{\mathbf{p}}
\newcommand{\bG}{\mathbf{G}}
\newcommand{\bR}{\mathbf{R}}
\newcommand{\bQ}{\mathbf{Q}}

\usepackage[toc,page]{appendix}
\newcommand{\nocontentsline}[3]{}
\newcommand{\tocless}[2]{\bgroup\let\addcontentsline=\nocontentsline#1{#2}\egroup}

\def\dfrac{\displaystyle\frac}

\makeatletter
\renewcommand{\p@subsection}{}
\renewcommand{\p@subsubsection}{}
\makeatother

\definecolor{darkerred}{rgb}{0,0,0}
\definecolor{blueish}{rgb}{0,0,0}

\begin{document}

\title{Exciton-exciton interactions in van der Waals heterobilayers}

\author{Alexander Steinhoff}
\email{asteinhoff@itp.uni-bremen.de}
\affiliation{Institut für Theoretische Physik, Universität Bremen, 28334 Bremen, Germany}
\affiliation{Bremen Center for Computational Materials Science, Universit\"at Bremen, 28334 Bremen, Germany}
\author{Edith Wietek}
\affiliation{Institute of Applied Physics and W\"urzburg-Dresden Cluster of Excellence ct.qmat, Technische Universit\"at Dresden, 01062 Dresden, Germany}
\author{Matthias Florian}
\affiliation{University of Michigan, Department of Electrical Engineering and Computer Science, Ann Arbor, Michigan 48109, USA}
\author{Tommy Schulz}
\affiliation{Institut für Theoretische Physik, Universität Bremen, 28334 Bremen, Germany}
\affiliation{Bremen Center for Computational Materials Science, Universit\"at Bremen, 28334 Bremen, Germany}
\author{Takashi Taniguchi}
\affiliation{Research Center for Materials Nanoarchitectonics, National Institute for Materials Science,  1-1 Namiki, Tsukuba 305-0044, Japan}
\author{Kenji Watanabe}
\affiliation{Research Center for Electronic and Optical Materials, National Institute for Materials Science, 1-1 Namiki, Tsukuba 305-0044, Japan}
\author{Shen Zhao}
\affiliation{Fakult\"at f\"ur Physik, Munich Quantum Center, and Center for NanoScience (CeNS), Ludwig-Maximilians-Universität M\"unchen, 80539 M\"unchen, Germany}
\author{Alexander Högele}
\affiliation{Fakult\"at f\"ur Physik, Munich Quantum Center, and Center for NanoScience (CeNS), Ludwig-Maximilians-Universität M\"unchen, 80539 M\"unchen, Germany}
\affiliation{Munich Center for Quantum Science and Technology (MCQST), 80799 M\"unchen, Germany}
\author{Frank Jahnke}
\affiliation{Institut für Theoretische Physik, Universität Bremen, 28334 Bremen, Germany}
\affiliation{Bremen Center for Computational Materials Science, Universit\"at Bremen, 28334 Bremen, Germany}
\author{Alexey Chernikov}
\affiliation{Institute of Applied Physics and W\"urzburg-Dresden Cluster of Excellence ct.qmat, Technische Universit\"at Dresden, 01062 Dresden, Germany}

\begin{abstract}
Exciton-exciton interactions are key to understanding non-linear optical and transport phenomena in van der Waals heterobilayers, which emerged as versatile platforms to study correlated electronic states.
We present a combined theory-experiment study of excitonic many-body effects based on first-principle band structures and Coulomb interaction matrix elements.
Key to our approach is the explicit treatment of the fermionic substructure of excitons and dynamical screening effects for density-induced energy renormalization and dissipation. 
We demonstrate that 
dipolar blue shifts are almost perfectly compensated by many-body effects, mainly by screening-induced self-energy corrections.
Moreover, we identify a crossover between attractive and repulsive behavior at elevated exciton densities. 
Theoretical findings are supported by experimental studies of spectrally-narrow interlayer excitons in atomically-reconstructed, hBN-encapsulated MoSe$_2$/WSe$_2$ heterobilayers.
Both theory and experiment show energy renormalization on a scale of a few meV even for high injection densities in the vicinity of the Mott transition.
Our results revise the established picture of dipolar repulsion dominating exciton-exciton interactions in van der Waals heterostructures and open up opportunities for their external design.
\end{abstract}

\maketitle


\section{Introduction}

Vertically stacked van der Waals heterobilayers with type-II band alignment host layer-separated, Coulomb-correlated electron-hole pairs forming interlayer excitons (ILX) with binding energies of more than $100$ meV and lifetimes that are often drastically increased in comparison to excitons within a single layer \cite{Fang2014,rivera_observation_2015, Rivera2018, merkl_ultrafast_2019, ovesen_interlayer_2019, peimyoo_electrical_2021, Barre2022}. 
For a wide range of electron-hole densities below the excitonic Mott transition \cite{steinhoff_exciton_2017,Wang2019b}, ILXs constitute an interacting quantum gas with renormalized spectral properties. 
Exciton-exciton interactions are of major importance in the context of understanding on-site repulsion energies and emerging many-body states in moiré systems\,\cite{Morrow2021, Li2020, Regan2022, Guo2022, Sun2022}, excitonic transport \cite{Wang2021, yuan_twist-angle-dependent_2020,sun_excitonic_2022,erkensten_microscopic_2022,Choi2023,wietek_non-linear_2023}, and non-linear optical response \cite{Miller2017,nagler_interlayer_2017,Li2021}. 
Importantly, they lead to excitation-dependent shifts of the exciton resonance towards higher energies, in close analogy to the physics of coupled quantum well systems \cite{butov_condensation_1994, butov_towards_2002, ivanov_quantum_2002, schindler_analysis_2008, laikhtman_exciton_2009, lobanov_theory_2016}. 
These energy shifts are broadly used as key observables to assess the strength of the exciton-exciton coupling and determine exciton densities\,\cite{Jauregui2019,Unuchek2019}.

Conventionally, the interaction is described in the semiclassical picture of long-range repulsion between two dipoles representing spatially indirect excitons\,\cite{Lozovik1996,Butov1999}.
Corresponding theoretical descriptions range from basic capacitor models \cite{ivanov_quantum_2002, lobanov_theory_2016} to correlation-corrected formulas \cite{schindler_analysis_2008, laikhtman_exciton_2009}, microscopic theories beyond the effective bosonic picture \cite{may_many-body_1985, combescot_effective_2002}, and adaptations of classical approaches to describe localized moiré excitons\,\cite{sun_enhanced_2022, gotting_moire-bose-hubbard_2022, lohof_confined-state_2023}.
Nevertheless, the dominating picture of exciton-exciton interactions broadly used in the field of van der Waals heterostructures remains based on the dipole-dipole repulsion.
Only recently, predictions of additional contributions emerged, associated with exchange interaction and Pauli blockade from the underlying fermionic substructure\,\cite{erkensten_microscopic_2022,katzer_exciton-phonon-scattering_2023}. 
This state of affairs strongly suggests to test and revise the dipolar description of the exciton-exciton interaction in van der Waals heterostructures, motivating the development of a comprehensive theoretical approach and a well-defined experimental setting.

\begin{figure*}[ht]
\centering
\includegraphics[width=1.\textwidth]{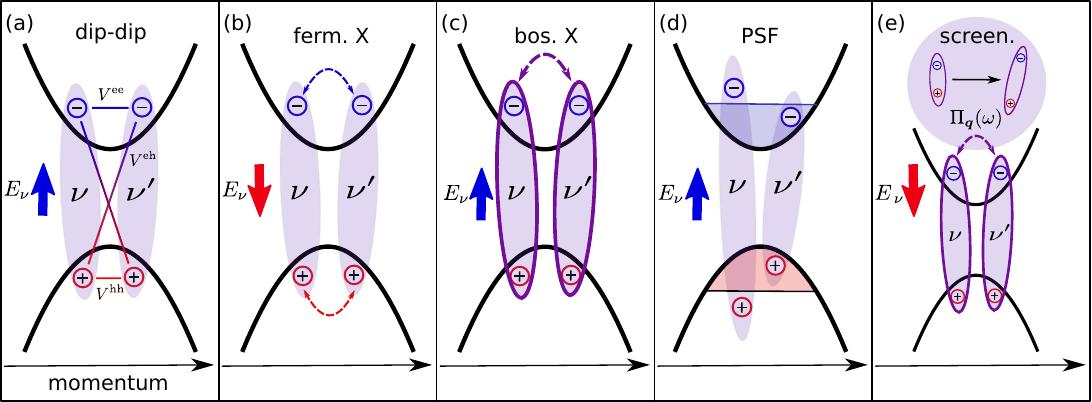}
\caption{\textbf{Contributions to exciton-exciton interaction.} (a) Dipole-dipole interaction (dip-dip) between excitons in states $\nu$ and $\nu'$. The process is composed of elementary Coulomb interaction between the electron of $\nu$ and the electron of $\nu'$,
proportional to the matrix element $V^{\textrm{ee}}$, between the electron of $\nu$ and the hole of $\nu'$ ($V^{\textrm{hh}}$) as well as corresponding electron-hole terms ($V^{\textrm{eh}}$). 
The resulting shift of the exciton energy $E_{\nu}$ is positive. 
(b) Fermionic exchange interaction (ferm. X) between electrons or holes that are part of two different excitons, resulting in a decrease of the exciton energy. 
(c) Bosonic exchange interaction (bos. X) of the whole excitons, resulting in an increase of the exciton energy. 
(d) Phase-space filling (PSF) due to fermionic constituents of exciton $\nu'$ as experienced by exciton $\nu$, which yields a blue shift of $E_{\nu}$.
(e) Screened bosonic exchange interaction (screen.) from the exciton Montroll-Ward (MW) self-energy. 
The interaction process is screened by a momentum- and frequency-dependent excitonic polarization $\Pi_{\bq}(\omega)$ corresponding to scattering processes of surrounding excitons between different internal states, which results in a lowering of exciton energy.}
\label{fig:X_X_interaction_full}
\end{figure*}

Here we present a combined theory-experiment study of density-dependent exciton energy renormalizations based on a systematic many-body description of interacting excitons as composite particles.
We determine both the resulting energy shifts and scattering-induced spectral broadening.
Supported by spectroscopic measurements of mobile excitons in atomically-reconstructed, hBN-encapsulated MoSe$_2$/WSe$_2$ heterobilayers, we find that the classical dipolar description drastically overestimates the interaction strength.
We show that the dipole-dipole repulsion is largely compensated by competing many-body effects within the dense exciton gas, including correction terms due to the fermionic substructure and screening. 
Several contributions have large absolute values, yet opposite signs.
As a net result, exciton energies shift by only a few meV even at exciton densities close to ionization threshold above $10^{12}$ cm$^{-2}$. 
Moreover, we demonstrate that the interplay of different renormalization effects can result in an effectively \textit{attractive} interaction with a density-dependent crossover to the repulsive regime.

To obtain quantitative results we consider a representative case of an H-stacked MoSe$_2$/WSe$_2$ heterobilayer encapsulated in hexagonal boron nitride (hBN) and build on a theory for the dense exciton gas based on two-particle Green functions \cite{may_many-body_1985, boldt_many-body_1985}. 
Key to our method development is the combination of this established many-body theory with material-realistic band structure and Coulomb matrix element calculations, augmented by frequency-dependent excitonic screening effects.
The starting point of our calculations is the Bethe-Salpeter equation in the absence of photoexcited carriers
\begin{equation}
 \begin{split}
 &(\varepsilon_{\bQ-\bk}^e+\varepsilon_{\bk}^h - E_{\nu,\bQ} )\Phi_{\nu,\bQ}(e,h,\bk) \\
 -&\sum_{\bq,h',e'} V^{e,h,h',e'}_{\bQ-\bk,\bk,\bk+\bq,\bQ-\bk-\bq} \Phi_{\nu,\bQ}(e',h',\bk+\bq)=0\,,
\end{split}
\label{eq:BSE_main}
\end{equation}
which we solve to obtain exciton wave functions $\Phi_{\nu,\bQ}(e,h,\bk)$ and energies $E_{\nu,\bQ}$. 
Here, exciton eigenstates are classified in terms of a quantum number $\nu$ for the relative electron-hole motion and the total exciton momentum $\bQ$. 
First-principle band structures $\varepsilon_{\bk}^a$ and dielectrically screened Coulomb matrix elements $V^{a,b,b',a'}_{\bk,\bk',\bk'+\bq,\bk-\bq}$ are used in the electron-hole picture, expanding beyond the macroscopic Coulomb interaction model and parabolic bands employed in the literature\,\cite{erkensten_microscopic_2022}.
In the exciton representation, we formulate a self-consistency equation for renormalized exciton energies as
\begin{equation}
 \begin{split}
\tilde{E}_{\nu,\bQ} & = E_{\nu,\bQ} + \textrm{Re}\, \Sigma(\nu,\bQ,\tilde{E}_{\nu,\bQ} / \hbar)\,,
\end{split}
\label{eq:Sigma_self_cons_main}
\end{equation}
with the frequency-dependent exciton self-energy
\begin{align}
\label{eq:self-energy}
\Sigma(\nu,\bQ,\omega) = & \Sigma^{\textrm{H}}(\nu,\bQ)+ \Sigma^{\textrm{F}}(\nu,\bQ)\\\notag
+ & \Sigma^{\textrm{PB}}(\nu,\bQ) + \Sigma^{\textrm{MW,ret}}(\nu,\bQ,\omega). 
\end{align}
Energy renormalizations are accompanied by excitation-induced broadening determined by $\Gamma_{\nu,\bQ}=-\textrm{Im}\, \Sigma(\nu,\bQ,\tilde{E}_{\nu,\bQ})$, where $2\Gamma_{\nu,\bQ}$ corresponds to the full-width-half-maximum (FWHM) of the homogeneous exciton linewidth.
The self-energy consists of Hartree (H), Fock (F), Pauli blocking (PB) and Montroll-Ward (MW) terms. 
Explicit expressions as well as a detailed derivation of the various self-energy contributions are given in the Appendix.
The different contributions to exciton-exciton interaction described by the exciton self-energy are schematically shown in Fig.~\ref{fig:X_X_interaction_full}.
A part of the Hartree term represents classical electrostatic interaction between excitons (Fig.~\ref{fig:X_X_interaction_full}(a)). 
It is of dipole-dipole type and contains two repulsive and two attractive contributions.
Since like charges reside within the same layer in type-II heterobilayers, the repulsive terms dominate over the attractive inter-layer terms. 
Therefore, the net effect is a pronounced blue shift of exciton energies.
On the other hand, exchange interaction between the fermionic constituents of excitons (included in both Hartree and Fock terms, Fig.~\ref{fig:X_X_interaction_full}(b)) leads to an attractive interaction as in a Fermi gas.
Excitons as composite bosons also experience a repulsive bosonic exchange interaction that is sketched in \ref{fig:X_X_interaction_full}(c), represented by the Fock term in Eq.\,\eqref{eq:self-energy}. 
This can be understood as the consequence of boson statistics favoring occupation of the same states that, in turn, leads to stronger Coulomb repulsion.

Pauli blocking of the fermionic phase space due to the exciton substructure is illustrated in Fig.\,\ref{fig:X_X_interaction_full}(d).
It acts as a source of the increased energy of the exciton transition, well-known from the semiconductor Bloch equations \cite{Haug2009, steinhoff_influence_2014, katzer_exciton-phonon-scattering_2023}. 
Finally, the Montroll-Ward self-energy contains all non-instantaneous contributions of the GW self-energy, describing frequency-dependent screening of bosonic exchange interaction by excitations within the dense exciton gas.
As discussed further below, this is a particularly important contribution that is explicitly treated beyond a more phenomenological description of static screening in the literature\,\cite{erkensten_microscopic_2022}.
The result is a decrease of the exciton energies, conceptually similar to the excitation-induced bandgap renormalization\,\cite{Haug2009,steinhoff_influence_2014,Chernikov2015,Ulstrup2016}.

\section{Experimental Setup}
The experiments were performed on a MoSe$_2$/WSe$_2$ heterobilayer encapsulated in high quality hBN. 
We employed fabrication conditions of near sixty-degree stacking favoring large-scale atomic reconstruction\,\cite{Zhao2023} suppressing formation of moiré patterns to achieve a flat potential landscape.
The H$_h^h$ registry was identified by its characteristic photoluminescence (PL) signatures, optical selection rules, and associated g-factors\,\cite{wietek_non-linear_2023}. 
The PL spectra feature low energy triplet- and high energy singlet-spin configuration with narrow linewidths of the exciton resonances down to 3 meV confirming the sample quality.
As localization of ILXs is suppressed in H$_h^h$ reconstructed samples, the excitons are mobile, exhibiting phonon-limited free diffusion at lowest temperatures.
Furthermore, signatures of exciton-exciton annihilation and repulsion validate that the excitons can scatter and interact efficiently with each other \cite{wietek_non-linear_2023}.

The measurements were performed in an optical microscopy cryostat at the temperatures of 5 and 70 K. 
The sample was excited by a pulsed, 140 fs Ti:sapphire laser with a repetition rate of 80\,MHz and its excitation wavelength tuned resonantly into the A:1s state of MoSe$_2$. 
The excitation energy densities ranged from 0.2 to 20 $\mu$J/cm$^{2}$. 
Assuming an absorbance of 11\%, as estimated from reflectance measurements, we determine the corresponding peak exciton densities of $6\times10^{10}$ and $6\times10^{12}$ cm$^{-2}$  (see Appendix for details).
The PL was spectrally dispersed by a spectrometer and detected by a streak camera and a charge-coupled device for time-resolved and time-integrated measurements, respectively.

\section{Results}

\begin{figure*}[ht]
\centering
\includegraphics[width=1.\textwidth]{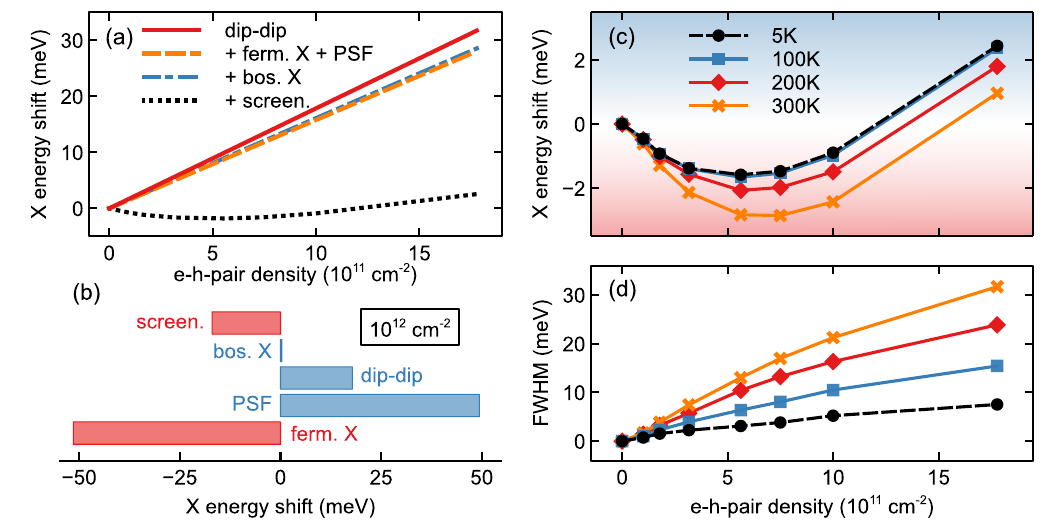}
\caption{\textbf{Exciton energy renormalization induced by exciton-exciton interaction.} 
(a) Cumulative density-dependent renormalization of the zero-momentum bright 1s-exciton (spin-singlet) energy at a temperature $T=100$ K, subsequently adding dipole-dipole interaction (dip-dip), fermionic exchange interaction and phase-space filling (+ ferm. X + PSF), bosonic exchange interaction (+ bos. X), and screened bosonic exchange (+ screen.). The latter represents the result of the full calculation.
(b) Individual contributions to the exciton energy renormalization corresponding to the cumulative presentation in panel (a) at an electron-hole pair density $10^{12}$ cm$^{-2}$. The red shift due to fermionic exchange and the blue shift due to phase-space filling compensate to a large extent.
(c) Calculated temperature and density dependence of energy renormalization for the zero-momentum bright 1s-exciton. 
The result for $T=5$ K has been obtained from extrapolating the high-temperature data. 
(d) Calculated temperature and density dependence of the zero-momentum bright 1s-exciton broadening.}
\label{fig:results}
\end{figure*}

To quantify many-body renormalizations due to the exciton-exciton interaction, we solve Eq.~(\ref{eq:Sigma_self_cons_main}) self-consistently for various exciton temperatures and densities. 
Dielectric screening by the hBN environment is modeled by an effectively isotropic dielectric constant $\varepsilon_{\textrm{hBN}} = \sqrt{4.95 \cdot 2.86}\approx 3.76$ \cite{artus_natural_2018}. 
We consider the inter-layer excitons populating the K- and K'-valleys of the two lowest conduction and two highest valence bands, respectively. 
The focus on K-valley states is motivated by their representative nature for interlayer excitons and by the absence of signatures from momentum-indirect excitons in the experiment\,\cite{wietek_non-linear_2023}. 
For the bright spin-singlet states we obtain a binding energy of $131$ meV in good agreement with typical experimental values \cite{merkl_ultrafast_2019}. 
Finally, we limit the exciton density range to be below the predicted exciton Mott density $n_{\text{Mott}}\approx 2\times10^{12}$ cm$^{-2}$ for this material system \cite{wietek_non-linear_2023}.

Numerical results of our calculations are shown in Fig.~\ref{fig:results}. 
In Fig.~\ref{fig:results}(a) we analyze the cumulative effect of all contributions to exciton-exciton interaction that have been introduced in Fig.~\ref{fig:X_X_interaction_full} for the 1s-exciton with vanishing total momentum $\bQ=0$ and bright spin-singlet configuration. 
The individual contributions are presented in Fig.~\ref{fig:results}(b).
The repulsive dipole-dipole interaction given by the exciton Hartree self-energy without fermionic correction terms (\ref{eq:exciton_Hartree_final_direct}) yields a linear blue shift of up to $30$ meV at the highest density close to $2\times10^{12}$ cm$^{-2}$. 
The blue shift is reduced by the combination of fermionic exchange contributions (\ref{eq:exciton_Hartree_final_X}) and (\ref{eq:exciton_Fock_final_X}), which are part of Hartree and Fock self-energies, respectively, and Pauli blocking. The latter originates from the filling of phase space by the fermionic constituents of excitons, as described by the self-energy in Eq.~(\ref{eq:exciton_PB_final}). 
Fermionic exchange and Pauli blocking, although on the order of $50$ meV each at a density of $10^{12}$ cm$^{-2}$, compensate each other to a large extent.

Bosonic exchange of excitons as described by the Fock self-energy without fermionic terms (\ref{eq:exciton_Fock_final_direct}) only yields a weak additional blue shift on the few meV scale. 
The relative weakness compared to the direct exciton-exciton interaction can be partly understood from the dependence of the different self-energies on exciton populations. 
While the Hartree-like renormalization of a certain exciton state is approximately proportional to the total exciton density, 
$\Sigma^{\textrm{H,(D)}}(\nu_1,\bQ_1)\approx \tilde{V}^{\textrm{(D)}}_{\bQ=0} \sum_{\nu'_1,\bQ'_1}N^{\textrm{X}}_{\nu'_1,\bQ'_1} $, the Fock-like term has a strong dependence on the momentum and spin distribution of excitons, $\Sigma^{\textrm{F,(D)}}(\nu_1,\bQ_1)\approx \sum_{\nu'_1,\bQ'_1} \tilde{V}^{\textrm{(D)}}_{|\bQ_1-\bQ'_1|}\delta_{s_e(\nu_1),s_e(\nu'_1)}\delta_{s_h(\nu_1),s_h(\nu'_1)} N^{\textrm{X}}_{\nu'_1,\bQ'_1} $.
Thus, bosonic exchange is sensitive only to a fraction of the total exciton density, which splits into intra- and inter-valley excitons with like and unlike electron $(s_e)$ and hole spins $(s_h)$. 


Finally, dynamical screening of the Coulomb interaction due to the presence of a polarizable exciton gas provides a substantial contribution to the energy renormalization.
It is represented by the Montroll-Ward self-energy (\ref{eq:MW_exciton}), which contains the non-instantaneous contributions to the exciton GW self-energy, leading to a red shift on the order of tens of meV. Due to a sub-linear increase of the red shift with the density, the cumulative energy renormalization is non-monotonous. 
Although this self-energy is determined by frequency-dependent screening caused by the polarization of the dense exciton gas, we can understand its attractive character in the static limit (\ref{eq:Sigma_SXCH}). 
Here, the self-energy splits into a contribution that leads to static screening of exciton exchange (without fermionic corrections) and a Coulomb hole (CH) contribution. 
The former yields a reduction of the weak momentum-dependent blue shift discussed above, while the latter is essentially a rigid red shift of the exciton dispersion. 
Since the CH shift is the dominant contribution, the Montroll-Ward self-energy is more sensitive to the total exciton density than to the distribution of excitons over the states.

\begin{figure*}[ht]
	\centering
	\includegraphics[width=1.\textwidth]{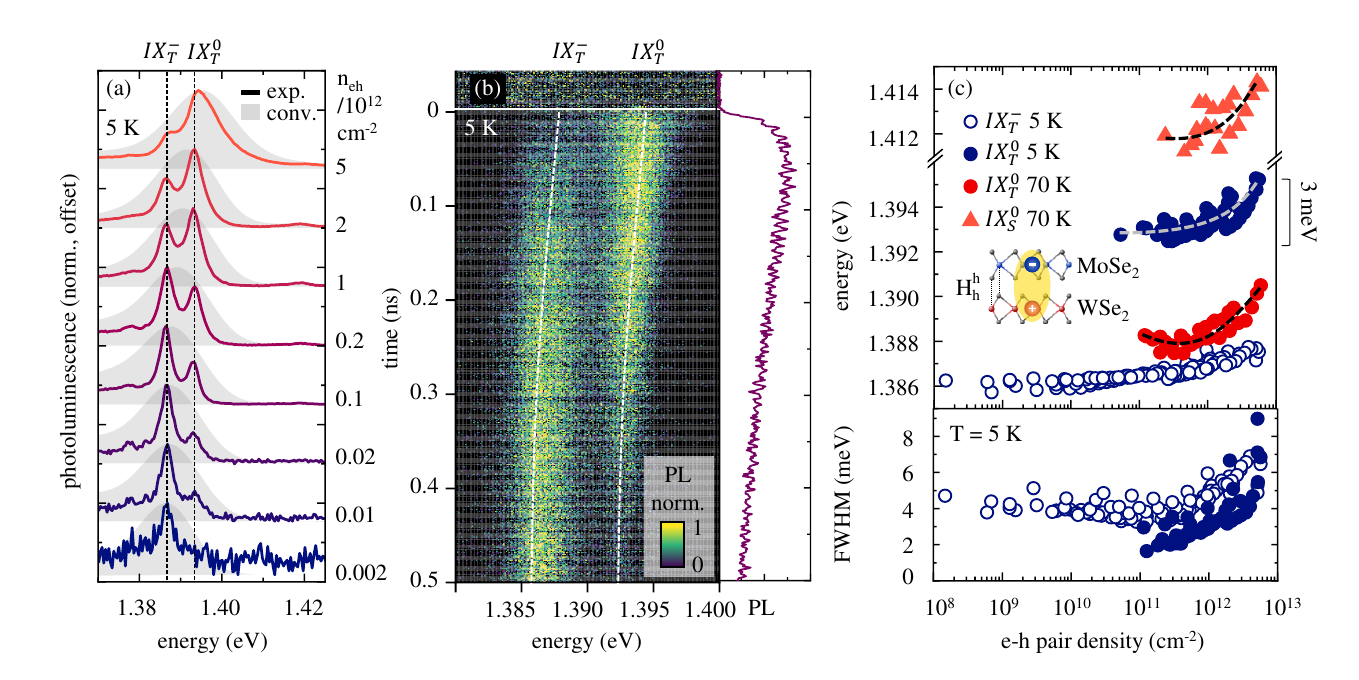}
	\caption{\textbf{Experimental energy shifts of interlayer excitons.}
	(a) Time integrated spectra of K-K’ interlayer excitons in an H$_h^h$ reconstructed  MoSe$_2$/WSe$_2$ heterobilayer for different exciton peak densities after pulsed excitation resonant with the MoSe$_2$ intralayer resonance at the lattice temperature of 5 K.
	For comparison, spectra convoluted with a Gaussian (full-width-at-half-maximum of 15 meV) are presented to simulate additional inhomogeneous broadening. 
	(b) Streak image of the spectrally- and time-resolved PL at a peak density of $10^{12}$ cm$^{-2}$. 
	The spectra are normalized at each time step and the color scale is logarithmic. White dashed lines are guides-to-the-eye, following the shift of the peak energy with time. 
	Corresponding PL transient is presented in the right panel. 
	(c) (upper panel) Density dependent energy shift of the interlayer exciton PL peaks at two different temperatures, extracted from time resolved data. 
	The resonances are labeled according to the previously identified excitons in the H$_h^h$-reconstructed MoSe$_2$/WSe$_2$ heterobilayer: charged and neutral interlayer triplets (IX$_T^0$ and IX$_T^-$) and the neutral interlayer singlet (IX$_S^0$). 
	Dashed lines are guides-to-the-eye.
	(lower panel) Corresponding density dependent linewidths of charged and neutral interlayer triplets at T = 5 K.}
	\label{fig:expresults}
\end{figure*}

To summarize these results in Fig.~\ref{fig:results}(a) and (b), the main competition takes place between the blue shift due to dipolar interaction and the red shift induced by excitonic screening, which are both sensitive to the total exciton density. 
This leads to the conclusion that the details of the exciton band structure are likely to be secondary for the total exciton shift. 
All in all, we find a remarkably strong compensation of different renormalization effects, leaving a net exciton blue shift of only a few meV even at high densities, close to the Mott transition.
This density-dependent behavior is consistently found both at low and high temperatures, as shown in Fig.~\ref{fig:results}(c) for spin-bright excitons. 
For all temperatures we find a few meV red shift at small and intermediate densities which turns into a few meV blue shift at high densities. 
This corresponds to a density-dependent crossover from an effectively attractive to repulsive exciton-exciton interaction.
With decreasing temperature, the attractive character of interaction becomes slightly weaker, yet also closer to the thermal energies of the excitons.
Moreover, we confirm that similar behavior is expected for ``grey'' (spin-triplet) inter-layer excitons, presented in Fig.~\ref{fig:results_dark} in the Appendix. 
Since the triplet exciton state is about $20$\,meV below the bright one, it is more strongly populated, which leads to an increase of fermionic and bosonic exchange effects among exciton triplets. 
As a result, the bright exciton interaction is slightly more repulsive.

Importantly, the non-instantaneous nature of the Montroll-Ward self-energy also results in a lifetime broadening of excitonic resonances according to Eq.~(\ref{eq:Sigma_self_cons}). 
Temperature- and density-dependent results are shown in Fig.~\ref{fig:results}(d). 
Overall, we find a sub-linear increase of broadening with the exciton density that is more pronounced at higher temperatures.
At $T=5$ K, we find a FWHM broadening of about $8$ meV at highest densities, slightly larger than the typical linewidths of the exciton resonances in these systems, likely dominated by inhomogeneous broadening.
For higher temperatures, however, exciton-exciton scattering can become comparable to the phonon-induced broadening at elevated densities, thus providing a substantial contribution to dissipation and dephasing.

Theoretical results are confirmed by experimental observations presented in Fig.~\ref{fig:expresults}. 
A selection of time integrated PL spectra for different exciton peak densities at a temperature of 5 K is shown in Fig.~\ref{fig:expresults}(a).
The narrow linewidths allow for the distinct assignment of the PL peaks of charged (IX$^-_T$) and neutral (IX$^0_T$) interlayer triplets of H$^h_h$ registry\,\cite{Zhao2023,wietek_non-linear_2023}. 
A small blue shift of only a few meV is observed for both exciton species up to highest densities in the range of the Mott transition. 
Simultaneously the relative PL weight shifts from IX$^-_T$ to IX$^0_T$ with increasing density.
We note, that for sufficiently large spectral broadening, this shift of the center of gravity would lead to the appearance of a much larger blue shift. 
To illustrate that, the spectra are convoluted with a Gaussian (full-width-at-half-maximum of 15 meV), so that the PL of charged and neutral exciton merges into one peak. 
By concealing the multi-peak structure of the emission, the density dependent blue shift is overestimated in contrast to weak shifts obtained for the individual peaks.

In addition to analyzing time-integrated spectra we use an alternative method of investigating the effect of density on ILX via time-resolved PL. 
Taking advantage of the exciton decay, we relate the time axis to the relative change in exciton density, gaining access to quasi-instantaneous measurements of the energy shift at a given density.
A representative time resolved PL evolution at a nominal temperature of 5 K and peak electron-hole pair density of $10^{12}$ cm$^{-2}$ is shown in Fig.~\ref{fig:expresults}(b). 
At time $t=0$ the injected electron-hole pair density is set equal to the estimated peak density. 
Consequently, at later times the exciton density decreases proportionally to the decaying PL intensity. 
This is reflected in the shift of the peak positions towards lower energies with time, as indicated by the dashed lines in Fig.~\ref{fig:expresults}(b). 
We evaluate the energy peak position for time steps of 50\,ps width and assign the corresponding densities to values given by the injected maximum density and the relative change of PL intensity.

The results of both time-integrated and time-resolved analysis are summarized in Fig.~\ref{fig:expresults}(c) upper panel for all measured triplet and singlet interlayer exciton states. 
The experiments were performed at a temperature of 5\,K to take advantage of the narrow linewidths and maximum PL yield, and at 70\,K, where the influence of the interlayer trion in PL is negligible. 
The maximum blue shift is found to be less or equal to 3\,meV for all studied exciton species in good agreement with theory.  
Weak indications of a red shift are observed for the triplet state at 70\,K.
The energy shifts are accompanied by a weak spectral broadening at higher densities, similar to theoretically predicted values, as illustrated in the lower panel of Fig.~\ref{fig:expresults}(c).
We also note, that the above observations are typical for samples with sufficiently narrow linewidths, as shown in the Appendix.
Overall, the experiments consistently reveal small energy shifts in the meV range, independent of the presence of charged excitons, observed at both low and elevated temperatures.

\section{Conclusions}

In conclusion, we have demonstrated that the established picture of dipolar repulsion 
of interlayer excitons is insufficient and severely overestimates energy shifts resulting from exciton-exciton interaction.
Among relevant contributions, we identify fermionic and bosonic exchange effects as well as the phase-space filling and excitonic screening that lead to renormalizations of comparable magnitude, but opposite signs. 
The main competition takes place between the repulsive dipolar interaction and the attractive screening-induced self-energy correction, which both essentially depend on the total exciton density. 
The net result is an energy shift of only a few meV at exciton densities up to the Mott transition.
Moreover, we demonstrate conditions for the emergence of weakly attractive exciton potentials, predicted to occur at low and intermediate densities. 
Interestingly, elevated temperatures yield a stronger tendency towards the attractive regime. 
In addition, the energy shifts are accompanied by scattering-induced spectral broadening of exciton states with a linear temperature dependence.
The theoretical predictions are confirmed in experiment by taking advantage of well-defined, atomically-reconstructed MoSe$_2$/WSe$_2$ heterobilayers void of localization with spectrally narrow resonances and mobile excitons.
The magnitude of the measured energy shifts agrees with the results of the theoretical calculations within the experimental uncertainty.
This holds both for low and elevated temperatures and is independent of the presence or absence of residual doping. 

The developed understanding of the excitonic interactions has wide implications for the interpretation of optical and nonlinear transport phenomena as well as the overall phase diagram of interlayer excitons. 
It challenges the common notion of dominant dipole-dipole repulsion and offers a more nuanced approach to understanding exciton-exciton interactions in van der Waals heterostructures.
Rendering the extraction of interlayer exciton densities from the energy shifts via capacitor model less useful than previously assumed, it highlights the importance of spectral analysis for samples featuring multiple, closely spaced resonances.
Interestingly, one can expect that individual contributions to the exciton energy renormalization could be tunable using distinct sample geometries, dielectric environments, and external fields. 
Making use of this tunability, it should even be possible to design and switch between effectively repulsive and attractive interaction regimes.
The implications range from the possible realization of local compression and excitonic droplets, to potentially favorable conditions for the formation of macroscopic many-body states such as superfluids and condensates.
Natural extensions of the presented approach and results towards interlayer excitons in moiré and moiré-like superlattices would have major consequences for the on-site interaction terms determining the correlations.
Overall, exciton-exciton interactions beyond the dipolar regime should have a substantial impact on the behavior of dense excitonic quantum gases on a fundamental level and be highly relevant for exploiting non-linearities in photonic and optoelectronic applications.

\clearpage
\newpage

\begin{widetext}

\section{Appendix}

\subsection{Theory}

To obtain a material-realistic description of renormalization effects induced by exciton-exciton interactions, we combine first-principle band structures and Coulomb interaction matrix elements with a many-body theory for the dense exciton gas based on nonequilibrium Green functions. We essentially follow the derivation given by May et al. \cite{may_many-body_1985, boldt_many-body_1985}, extending it with respect to (i) the generality of matrix elements and (ii) the inclusion of frequency-dependent screening effects.
\subsubsection{Density functional theory calculations, spin-orbit coupling and Coulomb matrix elements}
Density functional theory (DFT) calculations for a freestanding H$^h_h$ MoSe$_2$/WSe$_2$ heterobilayer are carried out using QUANTUM ESPRESSO V.6.6 \cite{giannozzi_quantum_2009, giannozzi_advanced_2017}. We apply the generalized gradient approximation (GGA) by Perdew, Burke, and Ernzerhof (PBE) \cite{perdew_generalized_1997} and use optimized norm-conserving Vanderbilt pseudopotential~\cite{van_setten_pseudodojo_2018} at a plane-wave cutoff of $80$~Ry. Uniform meshes (including the $\Gamma$-point) with $18\times18\times1$ k-points are combined with a Fermi-Dirac smearing of $5$~mRy. 
Using a fixed lattice constant of $a=3.29$~\AA\, \cite{gillen_interlayer_2018} and a fixed cell height of $35$~\AA, forces are minimized below $5\cdot 10^{-3}$~eV/\AA. The D3 Grimme method~\cite{grimme_consistent_2010} is used to include van-der-Waals corrections.
\\We use RESPACK \cite{nakamura_respack_2021} to construct a lattice Hamiltonian $H_0(\bk)$ in a 22-dimensional localized basis of Wannier orbitals (d$_{z^2}$, d$_{xz}$, d$_{yz}$, d$_{x^2-y^2}$ and d$_{xy}$ for Mo and W, respectively, p$_x$, p$_y$ and p$_z$ for Se) from the DFT results. 
We also calculate the dielectric function as well as bare and screened Coulomb matrix elements in the localized basis.
\\Spin-orbit interaction is included using an on-site $\boldsymbol{L\cdot S}$-coupling Hamiltonian, which is added to the 
non-relativistic Wannier Hamiltonian:
\begin{equation}
\begin{split}
H(\bk)=I_2 \otimes H_0(\bk)+H_{\textrm{SOC}}\,.
\end{split}
\label{eq:H_tot}
\end{equation}
Here, $I_2$ is the $2\times2$ identity matrix in the Hilbert space spanned by eigenstates $\ket{\uparrow}$ and $\ket{\downarrow}$ of the spin z component (perpendicular to the monolayer). We assume that the Coulomb matrix in Wannier representation is spin-independent and that spin-up and spin-down states are not mixed.
Diagonalization of $H(\bk)$ yields the band structure $\varepsilon_{\bk}^{\lambda}$ and the Bloch states $\ket{\psi_{\bk}^{\lambda}}=\sum_{\alpha} c^{\lambda}_{\alpha,\bk}\ket{\bk,\alpha} $,
where the coefficients $c^{\lambda}_{\alpha,\bk}$ describe the momentum-dependent contribution of the orbital $\alpha$ to the Bloch band $\lambda$.
The Bloch sums $\ket{\bk,\alpha}$ are connected to the localized basis via $\ket{\bk,\alpha}=\frac{1}{\sqrt{N}}\sum_{\bR}e^{i\bk\cdot\bR}\ket{\bR,\alpha}$ with the number of unit cells $N$ and lattice vectors $\bR$.
The SOC-Hamiltonian is given by 
\begin{equation}
\begin{split}
H_{\textrm{SOC}}=\frac{1}{\hbar^2}\boldsymbol{\tilde{L}\cdot S}=\frac{1}{2\hbar}\boldsymbol{\tilde{L}\cdot \sigma}
\end{split}
\label{eq:H_SOC}
\end{equation}
with the Pauli matrices $\boldsymbol{\sigma}=(\sigma_x,\sigma_y,\sigma_z)$ and the angular momentum operator provided in \cite{wietek_non-linear_2023}.
%
\\
\\Starting from the density-density-like bare Coulomb interaction matrix elements in the Wannier basis,
\begin{equation}
\begin{split}
U_{\alpha\beta}(\bq)=\sum_{\bR}e^{i\bq\cdot\bR}U_{\alpha\beta\beta\alpha}(\bR)
=\sum_{\bR}e^{i\bq\cdot\bR}\bra{\boldsymbol{0},\alpha}\bra{\bR,\beta} U(\br,\br') \ket{\bR,\beta}\ket{\boldsymbol{0},\alpha}\,,
\end{split}
\label{eq:U_1}
\end{equation}
and the corresponding (statically) screened matrix elements $V_{\alpha\beta}(\bq)$, we obtain an analytic description of Coulomb interaction in freestanding TMD heterobilayers that can be augmented by screening from a dielectric environment \cite{schonhoff_interplay_2016,steinhoff_exciton_2017, steinke_coulomb-engineered_2020}. 
The parametrization of Coulomb matrix elements for a freestanding H$^h_h$ MoSe$_2$/WSe$_2$ heterobilayer is provided in \cite{wietek_non-linear_2023}.
Environmental screening can be taken into account according to the Wannier function continuum electrostatics approach \cite{rosner_wannier_2015} that combines a macroscopic electrostatic model for the screening by the dielectric environment with a localized description of Coulomb interaction.
The macroscopic dielectric function of a freestanding bilayer embedded in a vertical heterostructure is obtained by solving Poisson's equation \cite{florian_dielectric_2018}.
\\We finally compute screened Coulomb matrix elements 
in the Bloch state representation by a unitary transformation using the coefficients $c^{\lambda}_{\alpha,\bk}$:
\begin{equation}
\begin{split}
V^{\lambda,\nu,\nu',\lambda'}_{\bk_1, \bk_2, \bk_3, \bk_4} = \frac{1}{\mathcal{A}}\sum_{\alpha, \beta} \big( c_{\alpha, \bk_1}^{\lambda} \big)^{*} \big( c_{\beta, \bk_2}^{\nu} \big)^{*} V^{\alpha\beta}_{\bk_1-\bk_4}c_{\beta, \bk_3}^{\nu'} c_{\alpha, \bk_4}^{\lambda'}  \, ,
\end{split}
\label{eq:Coul_ME}
\end{equation}
where $\bk_4=\bk_1+\bk_2-\bk_3+\bG$ due to momentum conservation and $\mathcal{A}$ denotes the crystal area.
\\Inspired by the discussion in \cite{gong_magnetoelectric_2013}, we assume that Bloch states are approximately spin-diagonal. We assign a definite spin to each band according to the dominant contribution given by the coefficients $c^{\lambda}_{\alpha,\bk}$. Furthermore, we make use of the fact that 
Coulomb interaction is spin-conserving, so that we can set Coulomb matrix elements $V^{\lambda,\nu,\nu',\lambda'}_{\bk_1 \bk_2 \bk_3 \bk_4}$ to zero if $\lambda$ and $\lambda'$ or $\nu$ and $\nu'$ belong to different spins.

\subsubsection{Derivation of a GW self-energy for excitons}
\begin{figure}
\centering
\includegraphics[width=.7\columnwidth]{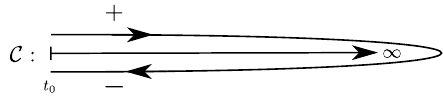}
\caption{The Keldysh contour $\mathcal{C}$ starts at initial time $t_0$, extends to $\infty$ on the upper branch ($+$) and returns to $t_0$ on the lower branch ($-$).}
\label{fig:keldysh_contour}
\end{figure}
We start from the definition of the exciton Schwinger-Keldysh Green function, 
\begin{equation}
\begin{split}
K(1,1')=\frac{1}{i\hbar}\ll h^{\phantom\dagger}_{\bk_1}(\underline{t}_1) e^{\phantom\dagger}_{\bp_1}(\underline{t}_1)e'^{\dagger}_{\bp'_1}(\underline{t}'_1)h'^{\dagger}_{\bk'_1}(\underline{t}'_1)\gg_{T} \, ,
\end{split}
\label{eq:exciton_GF}
\end{equation}
which describes the creation of an electron-hole pair at time $\underline{t}'_1$ and annihilation at time $\underline{t}_1$. 
The notation $\ll\cdot \gg_T$ is short for the statistical average
\begin{equation}
\begin{split}
\ll \hat{A}(\underline{t}) \gg_T=\frac{< T_{\mathcal{C}}\big[S_{\mathcal{C}} \hat{A}(\underline{t}) \big] >}{ <S_{\mathcal{C}}>}
\end{split}
\label{eq:stat_av}
\end{equation}
with the time evolution operator
\begin{equation}
\begin{split}
S_{\mathcal{C}}=T_{\mathcal{C}}\,\textrm{exp}\Big\{-\frac{i}{\hbar}\int_{\mathcal{C}} d\underline{t}'\,H_{\textrm{ext}}(\underline{t}') \Big\}
\end{split}
\label{eq:time_ev_op}
\end{equation}
and $T_{\mathcal{C}}$ being the time ordering operator on the Keldysh contour, which is depicted in Fig.~\ref{fig:keldysh_contour}. The Keldysh time coordinate $\underline{t}$ includes the branch index $n=\pm$ of the contour as well as the physical time coordinate $t$. 
The superindex $1$ stands for electron momentum $\bp_1$, hole momentum $\bk_1$, electron band $e$, hole band $h$ and Keldysh time $\underline{t}_1$.
The band index also contains the spin quantum number. In Eq.~(\ref{eq:exciton_GF}).
The external Hamiltonian
\begin{equation}
\begin{split}
H_{\textrm{ext}}(\underline{t})=\frac{e}{\mathcal{A}}\sum_{\bk,\bk',a,a'}V^{a',a}_{\textrm{ext}}(\bk,\bk',\underline{t})\rho^{a',a}(\bk',-\bk,\underline{t})
\end{split}
\label{eq:H_ext}
\end{equation}
couples the time-dependent external potential $V_{\textrm{ext}}$ to the particle-density operator
\begin{equation}
\begin{split}
\rho^{a',a}(\bk',\bk,\underline{t})=s_a a'^{\dagger}_{\bk'+\bk}(\underline{t}) a^{\phantom\dagger}_{\bk'}(\underline{t})\,.
\end{split}
\label{eq:dens_op}
\end{equation}
Here, we introduce the convention to assign a positive sign $s_a=+1$ to electrons ($a=e$) and a negative sign $s_a=-1$ to holes ($a=h$).
Whenever density-like operator pairs of the type in Eq.~(\ref{eq:dens_op}) appear, it is understood that both band indices describe the same carrier species.
\\In the following, we derive an equation of motion (EOM) for the exciton Green function (\ref{eq:exciton_GF}) by means of Heisenberg's equation for operators in the interaction picture with respect to $ H_{\textrm{ext}}(t)$,
\begin{equation}
\begin{split}
i\hbar\frac{\partial}{\partial t}\hat{A}(t)=\big[\hat{A}(t),H\big]\,,
\end{split}
\label{eq:heisenberg}
\end{equation}
where the Hamiltonian describes the interacting gas of electrons and holes:
\begin{equation}
\begin{split}
H=\sum_{\bp,e}\varepsilon_{\bp}^e e^{\dagger}_{\bp} e^{\phantom\dagger}_{\bp}+\sum_{\bk,h}\varepsilon_{\bk}^h h^{\dagger}_{\bk} h^{\phantom\dagger}_{\bk}
+&\frac{1}{2}\sum_{\bk,\bk',\bq}\sum_{e,e',a,a'} s_e s_a V^{e,a,a',e'}_{\bk,\bk',\bk'+\bq,\bk-\bq}e^{\dagger}_{\bk}a^{\dagger}_{\bk'} a^{\phantom\dagger}_{\bk'+\bq}e'^{\phantom\dagger}_{\bk-\bq}\\
+&\frac{1}{2}\sum_{\bk,\bk',\bq}\sum_{h,h',a,a'} s_h s_a V^{h,a,a',h'}_{\bk,\bk',\bk'+\bq,\bk-\bq}h^{\dagger}_{\bk}a^{\dagger}_{\bk'} a^{\phantom\dagger}_{\bk'+\bq}h'^{\phantom\dagger}_{\bk-\bq}\,.
\end{split}
\label{eq:H_electron_hole}
\end{equation}
Hole energies and wave functions are obtained from the valence-band quantities as $\varepsilon_{\bk}^h=-\varepsilon_{\bk}^v $ and $c_{\alpha, \bk}^{h}=\big( c_{\alpha, \bk}^{v} \big)^{*} $. 
The EOM for creation and annihilation operators can be derived using 
\begin{equation}
 \begin{split}
\left[ \hat{A}\hat{B}, \hat{C} \right]=\hat{A}\left[ \hat{B}, \hat{C} \right]_{+}-\left[ \hat{A}, \hat{C} \right]_+\hat{B}\,,
\end{split}
\label{eq:operator_identity}
\end{equation}
with the anti-commutator $\left[ \cdot\,,\cdot \right]_{+} $. To proceed, we apply the time-ordering operator in Eq.~(\ref{eq:exciton_GF}),
\begin{equation}
 \begin{split}
i\hbar\frac{\partial}{\partial t_1} K(1,1') = \frac{\partial}{\partial t_1}\Big\{ 
\theta_{\mathcal{C}}(\underline{t}_1,\underline{t}'_1) \ll h^{\phantom\dagger}_{\bk_1}(\underline{t}_1) e^{\phantom\dagger}_{\bp_1}(\underline{t}_1)e'^{\dagger}_{\bp'_1}(\underline{t}'_1)h'^{\dagger}_{\bk'_1}(\underline{t}'_1)\gg
+\theta_{\mathcal{C}}(\underline{t}'_1,\underline{t}_1) \ll h'^{\dagger}_{\bk'_1}(\underline{t}'_1)e'^{\dagger}_{\bp'_1}(\underline{t}'_1)e^{\phantom\dagger}_{\bp_1}(\underline{t}_1)h^{\phantom\dagger}_{\bk_1}(\underline{t}_1) \gg
\Big\}\,,
\end{split}
\label{eq:exciton_GF_time_ordering}
\end{equation}
where the step function on the Keldysh contour has been introduced as $\theta_{\mathcal{C}}(\underline{t}_1,\underline{t}_2)=\delta_{n_1,-}+n_1\delta_{n_1,n_2}\theta(t_1-t_2)$ to take all four combinations of branch indices $n_1$,$n_2$ into account correctly. We use brackets $\ll\cdot \gg$ to denote the statistical average (\ref{eq:stat_av}) without time ordering. The time derivative of expectation values is evaluated by splitting the exponential in $S_{\mathcal{C}}$ according to the Keldysh time ordering. Then we obtain
\begin{equation}
 \begin{split}
i\hbar\frac{\partial}{\partial t_1} K(1,1') &= n_1\delta_{n_1,n'_1}\delta(t_1-t'_1)F(1,1')+(\varepsilon_{\bp}^e+\varepsilon_{\bk}^h  ) K(1,1') \\
&+\frac{e}{\mathcal{A}}\sum_{\bk}\Big\{ \sum_{e''} V^{e, e''}_{\textrm{ext}}(\bp_1+\bk,\bk,\underline{t}_1)K(1,1')\Big |_{\bp_1+\bk,e''}  
- \sum_{h''} V^{h, h''}_{\textrm{ext}}(\bk_1+\bk,\bk,\underline{t}_1)K(1,1')\Big |_{\bk_1+\bk,h''}\Big\} \\
&-\sum_{\bq,h'',e''} V^{e,h,h'',e''}_{\bp_1,\bk_1,\bk_1+\bq,\bp_1-\bq} \,K(1,1')\Big |_{\bp_1-\bq,e''}^{\bk_1+\bq,h''}\\
&+\sum_{\bk,\bq,a,a',h''} s_h s_a V^{h,a,a',h''}_{\bk_1,\bk,\bk+\bq,\bk_1-\bq} \,r(1,1',\bk,\bq,a,a',\underline{t}_1^+)\Big|_{\bk_1-\bq,h''} \\
&+\sum_{\bk,\bq,a,a',e''} s_e s_a V^{e,a,a',e''}_{\bp_1,\bk,\bk+\bq,\bp_1-\bq} \,r(1,1',\bk,\bq,a,a',\underline{t}_1^+)\Big|_{\bp_1-\bq,e''}
   \,
\end{split}
\label{eq:exciton_GF_full_EOM}
\end{equation}
with the phase-space filling factor 
\begin{equation}
 \begin{split}
F(1,1')= \ll \hat{F}(1,1')\gg =\delta_{\bk_1,\bk'_1}\delta_{\bp_1,\bp'_1}\delta_{h,h'}\delta_{e,e'}
-\delta_{\bp_1,\bp'_1}\delta_{e,e'}\ll h'^{\dagger}_{\bk'_1}(\underline{t}_1) h^{\phantom\dagger}_{\bk_1}(\underline{t}_1)\gg-
\delta_{\bk_1,\bk'_1}\delta_{h,h'}\ll e'^{\dagger}_{\bp'_1}(\underline{t}_1) e^{\phantom\dagger}_{\bp_1}(\underline{t}_1) \gg
\end{split}
\label{eq:PSF_factor}
\end{equation}
and the three-particle Green function
\begin{equation}
 \begin{split}
 r(1,1',\bk,\bq,a,a',\underline{t})=
 \frac{1}{i\hbar}\ll a^{\dagger}_{\bk}(\underline{t}^+) a'^{\phantom\dagger}_{\bk+\bq}(\underline{t}) 
 h^{\phantom\dagger}_{\bk_1}(\underline{t}_1)e^{\phantom\dagger}_{\bp_1}(\underline{t}_1)e'^{\dagger}_{\bp'_1}(\underline{t}'_1)h'^{\dagger}_{\bk'_1}(\underline{t}'_1)\gg_T
   \,.
\end{split}
\label{eq:three_particle_GF}
\end{equation}
Here, $\underline{t}^+$ denotes a time that is infinitesimally later on the Keldysh contour than $\underline{t}$. We also introduced the notation 
$K(1,1')\Big |_{\bk_1+\bq,h''(\bp_1+\bq,e'')}$ for a Green function where the hole state (electron state) of argument $1$ is changed. By introducing the free inverse exciton Green function
\begin{equation}
 \begin{split}
 K^{-1}_0(1,1')=n_1\delta_{n_1,n'_1}\delta(t_1-t'_1)\Big\{ i\hbar\frac{\partial}{\partial t_1} \delta_{\bk_1,\bk'_1}\delta_{\bp_1,\bp'_1}\delta_{h,h'}\delta_{e,e'}
 -\mathcal{H}_0(1,1')-V_{\textrm{ext}}(1,1',\underline{t}_1)\Big\}
\end{split}
\label{eq:free_inverse_GF}
\end{equation}
with the effective exciton Hamiltonian
\begin{equation}
 \begin{split}
 \mathcal{H}_0(1,1')=(\varepsilon_{\bp}^e+\varepsilon_{\bk}^h  )\delta_{\bk_1,\bk'_1}\delta_{\bp_1,\bp'_1}\delta_{h,h'}\delta_{e,e'}
 -\sum_{\bq,h'',e''} V^{e,h,h'',e''}_{\bp_1,\bk_1,\bk_1+\bq,\bp_1-\bq} \delta_{\bk_1+\bq,\bk'_1}\delta_{\bp_1-\bq,\bp'_1}\delta_{h'',h'}\delta_{e'',e'}
\end{split}
\label{eq:exciton_H}
\end{equation}
and the external potential
\begin{equation}
 \begin{split}
 V_{\textrm{ext}}(1,1',\underline{t}_1)=\frac{e}{\mathcal{A}}\sum_{\bk}\Big\{ \sum_{e''} V^{e, e''}_{\textrm{ext}}(\bp_1+\bk,\bk,\underline{t}_1)\delta_{\bk_1,\bk'_1}\delta_{\bp_1+\bk,\bp'_1}\delta_{h,h'} \delta_{e'',e'}
- \sum_{h''} V^{h, h''}_{\textrm{ext}}(\bk_1+\bk,\bk,\underline{t}_1)
\delta_{\bk_1+\bk,\bk'_1}\delta_{\bp_1,\bp'_1}\delta_{h'',h'} \delta_{e,e'}
\Big\}
\end{split}
\label{eq:V_ext}
\end{equation}
we bring the EOM (\ref{eq:exciton_GF_full_EOM}) to its integral form:
\begin{equation}
 \begin{split}
\int_{\mathcal{C}}d\underline{t}_2 \sum_{\bk_2,\bp_2,h_2,e_2 }K^{-1}_0(1,2) K(2,1') &= n_1\delta_{n_1,n'_1}\delta(t_1-t'_1)F(1,1') \\
&+\sum_{\bk,\bq,a,a',h''} s_h s_a V^{h,a,a',h''}_{\bk_1,\bk,\bk+\bq,\bk_1-\bq} \,r(1,1',\bk,\bq,a,a',\underline{t}_1^+)\Big|_{\bk_1-\bq,h''} \\
&+\sum_{\bk,\bq,a,a',e''} s_e s_a V^{e,a,a',e''}_{\bp_1,\bk,\bk+\bq,\bp_1-\bq} \,r(1,1',\bk,\bq,a,a',\underline{t}_1^+)\Big|_{\bp_1-\bq,e''}
   \,.
\end{split}
\label{eq:exciton_GF_full_EOM_integral}
\end{equation}
The exciton Green function is coupled via Coulomb interaction to three-particle Green functions, which in turn are coupled to four-particle Green functions and so forth.
To truncate this hierarchy of EOM, we eliminate the three-particle Green function in Eq.~(\ref{eq:exciton_GF_full_EOM_integral}) by using the functional derivative with respect to the external potential, which is contained in the time evolution operator:
\begin{equation}
 \begin{split}
\frac{\delta K(1,1')}{\delta V^{b', b}_{\textrm{ext}}(\bq',\bq,\underline{t})}=\frac{e}{\mathcal{A}}\Big\{
s_b\frac{1}{i\hbar}r(1,1',\bq'-\bq,\bq,b',b,\underline{t})-K(1,1')d^{b',b}(\bq',-\bq,\underline{t})
\Big\}
\end{split}
\label{eq:func_deriv_K_Vext}
\end{equation}
with $d^{b',b}(\bq',-\bq,\underline{t})=\frac{1}{i\hbar}\ll \rho^{b',b}(\bq',-\bq,\underline{t}) \gg $. The integral EOM (\ref{eq:exciton_GF_full_EOM_integral}) is then written as
\begin{equation}
 \begin{split}
&\int_{\mathcal{C}}d\underline{t}_2 \sum_{\bk_2,\bp_2,h_2,e_2 }K^{-1}_{\textrm{eff}}(1,2) K(2,1') = n_1\delta_{n_1,n'_1}\delta(t_1-t'_1)F(1,1') \\
&+i\hbar\frac{\mathcal{A}}{e}\sum_{\bk,\bq,a,a'}\Big\{
\sum_{e''}V^{e,a,a',e''}_{\bp_1,\bk,\bk+\bq,\bp_1-\bq}\frac{\delta K(1,1')\Big|_{\bp_1-\bq,e''}}{\delta V^{a, a'}_{\textrm{ext}}(\bk+\bq,\bq,\underline{t}_1^+)}
-\sum_{h''}V^{h,a,a',h''}_{\bk_1,\bk,\bk+\bq,\bk_1-\bq}\frac{\delta K(1,1')\Big|_{\bk_1-\bq,h''}}{\delta V^{a, a'}_{\textrm{ext}}(\bk+\bq,\bq,\underline{t}_1^+)}
\Big\}
   \,.
\end{split}
\label{eq:exciton_GF_full_EOM_integral_func_deriv}
\end{equation}
The operator $ K^{-1}_{\textrm{eff}}(1,2)$ follows from $ K^{-1}_0(1,2)$ if the external potential $V_{\textrm{ext}}$ is replaced by the effective potential $V_{\textrm{eff}}$ that also contains the electrostatic potential of the mean charge density:
\begin{equation}
 \begin{split}
V^{a, a'}_{\textrm{eff}}(\bk+\bq,\bq,\underline{t}_1)=V^{a, a'}_{\textrm{ext}}(\bk+\bq,\bq,\underline{t}_1)
+\frac{\mathcal{A}}{e}\sum_{\bk',b,b'} V^{a,b,b',a'}_{\bk,\bk',\bk'-\bq,\bk+\bq}\rho^{b,b'}(\bk'-\bq,\bq,\underline{t}_1)
   \,.
\end{split}
\label{eq:V_eff_pot}
\end{equation}
We exploit the chain rule for functional derivatives to reformulate the interaction term on the RHS of (\ref{eq:exciton_GF_full_EOM_integral_func_deriv}):
\begin{equation}
 \begin{split}
\frac{\delta K(1,1')}{\delta V^{a, a'}_{\textrm{ext}}(\bk+\bq,\bq,\underline{t}_1^+)}=
\int_{\mathcal{C}}d\underline{t}_2\sum_{\bk_2,\bq_2,a_2,a'_2}
\frac{\delta K(1,1')}{\delta V^{a_2, a'_2}_{\textrm{eff}}(\bk_2+\bq_2,\bq_2,\underline{t}_2)}
\frac{\delta V^{a_2, a'_2}_{\textrm{eff}}(\bk_2+\bq_2,\bq_2,\underline{t}_2)}{\delta V^{a, a'}_{\textrm{ext}}(\bk+\bq,\bq,\underline{t}_1^+)}
   \,.
\end{split}
\label{eq:func_deriv_chain_rule}
\end{equation}
The second factor is identified as the inverse dielectric function
\begin{equation}
 \begin{split}
&\varepsilon^{-1}(\bk_2+\bq_2,\bq_2,a_2,a'_2,\underline{t}_2,\bk+\bq,\bq,a,a',\underline{t}_1)=\frac{\delta V^{a_2, a'_2}_{\textrm{eff}}(\bk_2+\bq_2,\bq_2,\underline{t}_2)}{\delta V^{a, a'}_{\textrm{ext}}(\bk+\bq,\bq,\underline{t}_1)}\\
=&n_1\delta_{n_1,n_2}\delta(t_1-t_2)\delta_{\bk,\bk_2}\delta_{\bq,\bq_2}\delta_{a,a_2}\delta_{a',a'_2}\\
+&\sum_{\bk',b,b'}V^{a_2,b,b',a'_2}_{\bk_2,\bk',\bk'-\bq_2,\bk_2+\bq_2}\Big\{
L(\bk'-\bq_2,\bq_2,b,b',\underline{t}_2,\bk+\bq,-\bq,a,a',\underline{t}_1)-i\hbar d^{b,b'}(\bk'-\bq_2,\bq_2,\underline{t}_2))d^{a,a'}(\bk+\bq,-\bq,\underline{t}_1)
\Big\}
\end{split}
\label{eq:inv_DF}
\end{equation}
with the density-density expectation value
\begin{equation}
 \begin{split}
L(\bk'-\bq_2,\bq_2,b,b',\underline{t}_2,\bk+\bq,-\bq,a,a',\underline{t}_1)=
\frac{1}{i\hbar}\ll
\rho^{a,a'}(\bk+\bq,-\bq,\underline{t}_1)\rho^{b,b'}(\bk'-\bq_2,\bq_2,\underline{t}_2)
\gg_T \,.
\end{split}
\label{eq:def_L}
\end{equation}
Also, by defining the inverse exciton Green function via
\begin{equation}
 \begin{split}
\int_{\mathcal{C}}d\underline{t}_2\sum_{\bk_2,\bp_2,h_2,e_2}K^{-1}(1,2)K(2,1')=n_1\delta_{n_1,n'_1}\delta(t_1-t'_1)\delta_{\bk_1,\bk'_1}\delta_{\bp_1,\bp'_1}\delta_{h,h'}\delta_{e,e'}
 \,,
\end{split}
\label{eq:inverse_X_GF}
\end{equation}
we find for the polarization function
\begin{equation}
 \begin{split}
\Pi(3,1',\bk_4+\bq_4,\bq_4,a_4, a'_4,\underline{t}_4)&=\frac{\delta K(3,1')}{\delta V^{a_4, a'_4}_{\textrm{eff}}(\bk_4+\bq_4,\bq_4,\underline{t}_4)} \\
&=\frac{e}{\mathcal{A}}\int_{\mathcal{C}} d\underline{t}_1 \int_{\mathcal{C}} d\underline{t}_2 
\sum_{\bk_1,\bp_1,h_1,e_1}\sum_{\bk_2,\bp_2,h_2,e_2}K(3,1) \Gamma(1,2,\bk_4+\bq_4,\bq_4,a_4,a'_4,\underline{t}_4) K(2,1')
\end{split}
\label{eq:deriv_K_Veff}
\end{equation}
with the vertex function
\begin{equation}
 \begin{split}
\Gamma(1,2,\bk_4+\bq_4,\bq_4,a_4,a'_4,\underline{t}_4)=-\frac{\mathcal{A}}{e}
\frac{\delta K^{-1}(1,2)}{\delta V^{a_4, a'_4}_{\textrm{eff}}(\bk_4+\bq_4,\bq_4,\underline{t}_4)}\,.
\end{split}
\label{eq:vertexfct}
\end{equation}
Combining Eqs.~(\ref{eq:exciton_GF_full_EOM_integral_func_deriv}), (\ref{eq:func_deriv_chain_rule}), (\ref{eq:inv_DF}) and (\ref{eq:deriv_K_Veff}), we
arrive at the Dyson equation for the exciton Green function:
\begin{equation}
 \begin{split}
&\int_{\mathcal{C}}d\underline{t}_2 \sum_{\bk_2,\bp_2,h_2,e_2 }K^{-1}_{\textrm{eff}}(1,2) K(2,1') = n_1\delta_{n_1,n'_1}\delta(t_1-t'_1)F(1,1') \\
&+i\hbar\sum_{\bk,\bq,a,a'}
\int_{\mathcal{C}} d\underline{t}_2\sum_{\bk_2,\bq_2,a_2,a'_2}
\varepsilon^{-1}(\bk_2+\bq_2,\bq_2,a_2,a'_2,\underline{t}_2,\bk+\bq,\bq,a,a',\underline{t}_1)\\
&\times\int_{\mathcal{C}} d\underline{t}_3 \sum_{\bk_3,\bp_3,h_3,e_3}
\Big\{
\sum_{e''}V^{e,a,a',e''}_{\bp_1,\bk,\bk+\bq,\bp_1-\bq}K(1,3)\Big|_{\bp_1-\bq,e''}
-\sum_{h''}V^{h,a,a',h''}_{\bk_1,\bk,\bk+\bq,\bk_1-\bq}K(1,3)\Big|_{\bk_1-\bq,h''}
\Big\}\\
&\times \int_{\mathcal{C}} d\underline{t}_4 
\sum_{\bk_4,\bp_4,h_4,e_4}
\Gamma(3,4,\bk_2+\bq_2,\bq_2,a_2,a'_2,\underline{t}_2)K(4,1')\\
&=n_1\delta_{n_1,n'_1}\delta(t_1-t'_1)F(1,1')+\int_{\mathcal{C}} d\underline{t}_4 
\sum_{\bk_4,\bp_4,h_4,e_4}\Sigma(1,4)K(4,1')
\end{split}
\label{eq:exciton_GF_Dyson_general}
\end{equation}
with the self-energy $ \Sigma(1,1')$. Eq.~(\ref{eq:exciton_GF_Dyson_general}) is an exact EOM for the exciton Green function, where the many-body interaction effects are contained in the self-energy, or, more specifically, in the inverse dielectric function and the vertex function. 
To solve the many-body problem for the dense exciton gas, approximations have to be applied to the self-energy.
\\We derive the lowest approximation to the vertex function (\ref{eq:vertexfct}) by neglecting the density terms in the phase-space filling factor $F(1,1')$ and inverting Eq.~(\ref{eq:exciton_GF_Dyson_general}) to obtain an EOM for the inverse exciton Green function:
\begin{equation}
 \begin{split}
K^{-1}(1,1') = K^{-1}_{\textrm{eff}}(1,1') - \Sigma(1,1')\,.
\end{split}
\label{eq:inverse_exciton_GF_Dyson}
\end{equation}
The derivative of $K^{-1}_{\textrm{eff}}(1,1')$ with respect to $V_{\textrm{eff}}$ then yields a delta-like vertex function:
\begin{equation}
 \begin{split}
\Gamma(3,4,\bk_2+\bq_2,\bq_2,a_2,a'_2,\underline{t}_2)\approx
&n_3\delta_{n_3,n_4}\delta(t_3-t_4)n_3\delta_{n_3,n_2}\delta(t_3-t_2)\\
\times&\Big\{
\delta_{\bk_3,\bk_4}\delta_{\bp_3+\bq_2,\bp_4}\delta_{\bk_2,\bp_3}\delta_{h_3,h_4}\delta_{a_2,e_3}\delta_{a'_2,e_4}
-\delta_{\bp_3,\bp_4}\delta_{\bk_3+\bq_2,\bk_4}\delta_{\bk_2,\bk_3}\delta_{e_3,e_4}\delta_{a_2,h_3}\delta_{a'_2,h_4}
\Big\}
\,.
\end{split}
\label{eq:vertexfct_approx}
\end{equation}
Inserting this into Eq.~(\ref{eq:exciton_GF_Dyson_general}), we obtain the self-energy in random-phase approximation (RPA):
\begin{equation}
 \begin{split}
\Sigma(1,1')=i\hbar \sum_{\bk,\bq,\bq_2,a,a'}\Big\{\sum_{\bar{e}}
&\varepsilon^{-1}(\bp'_1,\bq_2,\bar{e},e',\underline{t}'_1,\bk+\bq,\bq,a,a',\underline{t}_1^+)\\
\times &\Big(
\sum_{e''}V^{e,a,a',e''}_{\bp_1,\bk,\bk+\bq,\bp_1-\bq}K(1,1')\Big|_{\bp'_1-\bq_2,\bar{e}}^{\bp_1-\bq,e''}
-\sum_{h''}V^{h,a,a',h''}_{\bk_1,\bk,\bk+\bq,\bk_1-\bq}K(1,1')\Big|_{\bp'_1-\bq_2,\bar{e}}^{\bk_1-\bq,h''}
\Big)\\
+&\sum_{\bar{h}}\varepsilon^{-1}(\bk'_1,\bq_2,\bar{h},h',\underline{t}'_1,\bk+\bq,\bq,a,a',\underline{t}_1^+)\\
\times &\Big(
\sum_{h''}V^{h,a,a',h''}_{\bk_1,\bk,\bk+\bq,\bk_1-\bq}K(1,1')\Big|_{\bk'_1-\bq_2,\bar{h}}^{\bk_1-\bq,h''}
-\sum_{e''}V^{e,a,a',e''}_{\bp_1,\bk,\bk+\bq,\bp_1-\bq}K(1,1')\Big|_{\bk'_1-\bq_2,\bar{h}}^{\bp_1-\bq,e''}
\Big)
\Big\}\,.
\end{split}
\label{eq:Sigma_GW}
\end{equation}
The matrix elements $V$ contain background screening from the filled valence bands as well as from the dielectric environment.
On the other hand, the inverse dielectric function $\varepsilon^{-1}$ describes screening from photoexcited electrons and holes.
Therefore the (matrix) product $\varepsilon^{-1} V$ can be identified as the fully screened Coulomb interaction $W$, which gives the self-energy
(\ref{eq:Sigma_GW}) the well-known GW form. As main difference to the standard GW self-energy, its excitonic version contains 
four interaction terms between the electrons and holes constituting two interacting excitons. Note that Coulomb interaction between the electron
and hole within the same exciton is already included in the free inverse exciton Green function (\ref{eq:free_inverse_GF}).
However, as discussed in Ref.~\cite{may_many-body_1985}, the RPA self-energy does not describe effects due to the exchange of fermions \textit{between} two excitons.
In the following subsection, we again follow Ref.~\cite{may_many-body_1985} to derive such fermionic corrections to the GW self-energy. 

\subsubsection{Exchange corrections to the GW self-energy}

Instead of eliminating the three-particle Green function $r$ by means of functional derivatives from the EOM of
the exciton Green function (\ref{eq:exciton_GF_full_EOM}), we derive an EOM for $r$ itself using an alternative
decoupling mechanism. To this end, we make use of the free inverse exciton GF (\ref{eq:free_inverse_GF}):
\begin{equation}
 \begin{split}
 &\int_{\mathcal{C}}d\underline{t}_2 \sum_{\bk_2,\bp_2,h_2,e_2 }
 r(1,2,\bk,\bq,a,a',\underline{t}_1^+)
 K^{-1}_0(2,1')
 \\=
 &\int_{\mathcal{C}}d\underline{t}_2 \sum_{\bk_2,\bp_2,h_2,e_2 }
 r(1,2,\bk,\bq,a,a',\underline{t}_1^+)
 n_2\delta_{n_2,n'_1}\delta(t_2-t'_1)\Big\{ i\hbar\frac{\partial}{\partial t_2} \delta_{\bk_2,\bk'_1}\delta_{\bp_2,\bp'_1}\delta_{h_2,h'}\delta_{e_2,e'}
 -\mathcal{H}_0(2,1')-V_{\textrm{ext}}(2,1',\underline{t}_2)\Big\}\,.
\end{split}
\label{eq:EOM_r}
\end{equation}
Since $K^{-1}_0$ acts to the left, we have to apply the adjoint operators in each term. The time derivative
\begin{equation}
 \begin{split}
 -i\hbar\frac{\partial}{\partial t'_1} r(1,1',\bk,\bq,a,a',\underline{t}_1^+)
 =-\frac{\partial}{\partial t'_1}\Big\{&
 \theta_{\mathcal{C}}(\underline{t}_1,\underline{t}'_1)
 \ll a^{\dagger}_{\bk}(\underline{t}_1) a'^{\phantom\dagger}_{\bk+\bq}(\underline{t}_1) 
 h^{\phantom\dagger}_{\bk_1}(\underline{t}_1)e^{\phantom\dagger}_{\bp_1}(\underline{t}_1)e'^{\dagger}_{\bp'_1}(\underline{t}'_1)h'^{\dagger}_{\bk'_1}(\underline{t}'_1)\gg \\
 +&\theta_{\mathcal{C}}(\underline{t}'_1,\underline{t}_1)
 \ll e'^{\dagger}_{\bp'_1}(\underline{t}'_1) h'^{\dagger}_{\bk'_1}(\underline{t}'_1)  a^{\dagger}_{\bk}(\underline{t}_1) a'^{\phantom\dagger}_{\bk+\bq}(\underline{t}_1) 
 h^{\phantom\dagger}_{\bk_1}(\underline{t}_1)e^{\phantom\dagger}_{\bp_1}(\underline{t}_1)\gg\Big\}
 \end{split}
\label{eq:time_deriv_r}
\end{equation}
is evaluated along the same lines as for the two-particle Green function (\ref{eq:exciton_GF_time_ordering}). Note that in (\ref{eq:time_deriv_r}), the time derivative is acting on creation instead of annihilation operators. Multiplying Eq.~(\ref{eq:EOM_r}) with $K_0(1',3)$ from the right, integrating over the superindex $1'$ and using the
relation
\begin{equation}
 \begin{split}
\int_{\mathcal{C}}d\underline{t}_2\sum_{\bk_2,\bp_2,h_2,e_2}K_0^{-1}(1,2)K_0(2,1')=n_1\delta_{n_1,n'_1}\delta(t_1-t'_1)\delta_{\bk_1,\bk'_1}\delta_{\bp_1,\bp'_1}\delta_{h,h'}\delta_{e,e'}
 \,,
\end{split}
\label{eq:inverse_X_GF_0}
\end{equation}
we arrive at the following explicit expression for $r$:
\begin{equation}
 \begin{split}
 &r(1,3,\bk,\bq,a,a',\underline{t}_1^+)\\
 &=
 \int_{\mathcal{C}}d\underline{t}'_1\sum_{\bk'_1,\bp'_1,h',e'}
 r_0(1,1',\bk,\bq,a,a',\underline{t}_1^+)K_0(1',3) \\
 &+\int_{\mathcal{C}}d\underline{t}'_1\sum_{\bk'_1,\bp'_1,h',e'}\sum_{\bk',\bq',a'',a'''}\\
 \times
 \Big\{&\sum_{h''}s_{h'}s_{a''}V^{h'',a'',a''',h'}_{\bk'_1+\bq',\bk',\bk'+\bq',\bk'_1}
 s(1,1',\bk,\bq,a,a',\underline{t}_1,\bk',\bq',a'',a''',\underline{t}'_1)\Big|_{\bk'_1+\bq',h''} \\ 
 &\sum_{e''}s_{e'}s_{a''}V^{e'',a'',a''',e'}_{\bp'_1+\bq',\bk',\bk'+\bq',\bp'_1}
 s(1,1',\bk,\bq,a,a',\underline{t}_1,\bk',\bq',a'',a''',\underline{t}'_1)\Big|_{\bp'_1+\bq',e''}
 \Big\}K_0(1',3)
 \,.
\end{split}
\label{eq:r_explicit}
\end{equation}
Here, $r_0$ results from the time derivative of the theta functions in Eq.~(\ref{eq:time_deriv_r}), for which we find
\begin{equation}
 \begin{split}
 r_0(1,1',\bk,\bq,a,a',\underline{t}_1^+)&=n_1\delta_{n_1,n'_1}\delta(t_1-t'_1)\ll\left[
 a^{\dagger}_{\bk}(\underline{t}_1) a'^{\phantom\dagger}_{\bk+\bq}(\underline{t}_1) 
 h^{\phantom\dagger}_{\bk_1}(\underline{t}_1)e^{\phantom\dagger}_{\bp_1}(\underline{t}_1),
 e'^{\dagger}_{\bp'_1}(\underline{t}_1)h'^{\dagger}_{\bk'_1}(\underline{t}_1)
 \right]\gg \\
 &=n_1\delta_{n_1,n'_1}\delta(t_1-t'_1)\Big\{ 
 \ll a^{\dagger}_{\bk}(\underline{t}_1) a'^{\phantom\dagger}_{\bk+\bq}(\underline{t}_1) \hat{F}(1,1')  \gg \\
 &+\delta_{\bp'_1,\bk+\bq}\delta_{e',a'}\ll a^{\dagger}_{\bk}(\underline{t}_1)h'^{\dagger}_{\bk'_1}(\underline{t}_1)h^{\phantom\dagger}_{\bk_1}(\underline{t}_1)e^{\phantom\dagger}_{\bp_1}(\underline{t}_1) \gg
 +\delta_{\bk'_1,\bk+\bq}\delta_{h',a'}\ll e'^{\dagger}_{\bp'_1}(\underline{t}_1)a^{\dagger}_{\bk}(\underline{t}_1)h^{\phantom\dagger}_{\bk_1}(\underline{t}_1)e^{\phantom\dagger}_{\bp_1}(\underline{t}_1) \gg
 \Big\}
 \,.
\end{split}
\label{eq:r0_explicit}
\end{equation}
The three-particle Green function couples to four-particle Green functions of the type
\begin{equation}
 \begin{split}
 s(1,1',\bk,\bq,a,a',\underline{t},\bk',\bq',a'',a''',\underline{t}')=
 \frac{1}{i\hbar}\ll a^{\dagger}_{\bk}(\underline{t}^+) a'^{\phantom\dagger}_{\bk+\bq}(\underline{t}) 
 h^{\phantom\dagger}_{\bk_1}(\underline{t}_1)e^{\phantom\dagger}_{\bp_1}(\underline{t}_1)e'^{\dagger}_{\bp'_1}(\underline{t}'_1)h'^{\dagger}_{\bk'_1}(\underline{t}'_1)
 a''^{\dagger}_{\bk'}(\underline{t}'^+) a'''^{\phantom\dagger}_{\bk'+\bq'}(\underline{t}') \gg_T
   \,.
\end{split}
\label{eq:four_particle_GF}
\end{equation}
The expression (\ref{eq:r_explicit}) is inserted in the EOM for the two-particle Green function (\ref{eq:exciton_GF_full_EOM_integral}), which yields
\begin{equation}
 \begin{split}
\int_{\mathcal{C}}d\underline{t}_2 \sum_{\bk_2,\bp_2,h_2,e_2 }K^{-1}_0(1,2) K(2,1') = n_1\delta_{n_1,n'_1}\delta(t_1-t'_1)F(1,1')
+\int_{\mathcal{C}}d\underline{t}_2 \sum_{\bk_2,\bp_2,h_2,e_2 } M(1,2) K_0(2,1')
\end{split}
\label{eq:exciton_GF_full_EOM_integral_alternative}
\end{equation}
with 
\begin{equation}
 \begin{split}
M(1,1')=\sum_{\bk,\bq,a,a'}\Bigg\{
\sum_{\bar{h}}s_a s_h V^{h,a,a',\bar{h}}_{\bk_1,\bk,\bk+\bq,\bk_1-\bq}\Bigg(&r_0(1,1',\bk,\bq,a,a',\underline{t}_1^+)\Big|_{\bk_1-\bq,\bar{h}} \\
+\sum_{\bk',\bq',a'',a'''}\Big\{&\sum_{h''}s_{h'}s_{a''}V^{h'',a'',a''',h'}_{\bk'_1+\bq',\bk',\bk'+\bq',\bk'_1}
s(1,1',\bk,\bq,a,a',\underline{t}_1,\bk',\bq',a'',a''',\underline{t}'_1)\Big|_{\bk'_1+\bq',h''}^{\bk_1-\bq,\bar{h}}\\
+&\sum_{e''}s_{e'}s_{a''}V^{e'',a'',a''',e'}_{\bp'_1+\bq',\bk',\bk'+\bq',\bp'_1}
s(1,1',\bk,\bq,a,a',\underline{t}_1,\bk',\bq',a'',a''',\underline{t}'_1)\Big|_{\bp'_1+\bq',e''}^{\bk_1-\bq,\bar{h}}\Big\}\Bigg) \\
+\sum_{\bar{e}}s_a s_e V^{e,a,a',\bar{e}}_{\bp_1,\bk,\bk+\bq,\bp_1-\bq}\Bigg(&r_0(1,1',\bk,\bq,a,a',\underline{t}_1^+)\Big|_{\bp_1-\bq,\bar{e}} \\
+\sum_{\bk',\bq',a'',a'''}\Big\{&\sum_{h''}s_{h'}s_{a''}V^{h'',a'',a''',h'}_{\bk'_1+\bq',\bk',\bk'+\bq',\bk'_1}
s(1,1',\bk,\bq,a,a',\underline{t}_1,\bk',\bq',a'',a''',\underline{t}'_1)\Big|_{\bk'_1+\bq',h''}^{\bp_1-\bq,\bar{e}}\\
+&\sum_{e''}s_{e'}s_{a''}V^{e'',a'',a''',e'}_{\bp'_1+\bq',\bk',\bk'+\bq',\bp'_1}
s(1,1',\bk,\bq,a,a',\underline{t}_1,\bk',\bq',a'',a''',\underline{t}'_1)\Big|_{\bp'_1+\bq',e''}^{\bp_1-\bq,\bar{e}}\Big\}\Bigg)\Bigg\}
\end{split}
\label{eq:alternative_self_energy}
\end{equation}
Eq.~(\ref{eq:exciton_GF_full_EOM_integral_alternative}) is not a full Dyson equation, since only the free exciton Green function $K_0(2,1')$ appears on the RHS. It rather represents the first skeleton diagram of an expansion of the full exciton Green function with respect to $M$. As discussed in Ref.~\cite{may_many-body_1985}, it is meaningful to interpret $M$ as a self-energy and upgrade $K_0(2,1')$ to $K(2,1')$. Indeed, it turns out that the four-particle Green functions can be factorized into two-particle Green functions such that the GW self-energy obtained from functional derivatives is reproduced. For example, the first four-particle term in Eq.~(\ref{eq:alternative_self_energy}) can be rewritten as:
\begin{equation}
 \begin{split}
&\sum_{\bk,\bq,a,a'}\sum_{\bk',\bq',a'',a'''}\sum_{\bar{h}}\sum_{h''}s_a s_{a''} V^{h,a,a',\bar{h}}_{\bk_1,\bk,\bk+\bq,\bk_1-\bq}V^{h'',a'',a''',h'}_{\bk'_1+\bq',\bk',\bk'+\bq',\bk'_1}
s(1,1',\bk,\bq,a,a',\underline{t}_1,\bk',\bq',a'',a''',\underline{t}'_1)\Big|_{\bk'_1+\bq',h''}^{\bk_1-\bq,\bar{h}} \\
=\frac{1}{i\hbar}&\sum_{\bk,\bq,a,a'}\sum_{\bk',\bq',a'',a'''}\sum_{\bar{h}}\sum_{h''}s_a s_{a''} V^{h,a,a',\bar{h}}_{\bk_1,\bk,\bk+\bq,\bk_1-\bq}V^{h'',a'',a''',h'}_{\bk'_1+\bq',\bk',\bk'+\bq',\bk'_1}\\
&\times\ll 
a^{\dagger}_{\bk}(\underline{t}_1^+) a'^{\phantom\dagger}_{\bk+\bq}(\underline{t}_1) 
 \bar{h}^{\phantom\dagger}_{\bk_1-\bq}(\underline{t}_1)e^{\phantom\dagger}_{\bp_1}(\underline{t}_1)
 e'^{\dagger}_{\bp'_1}(\underline{t}'_1)h''^{\dagger}_{\bk'_1+\bq'}(\underline{t}'_1)
 a''^{\dagger}_{\bk'}
 (\underline{t}'^+_1) 
a'''^{\phantom\dagger}_{\bk'+\bq'}(\underline{t}'_1) \gg_T \\
\approx\frac{1}{i\hbar}&\sum_{\bk,\bq,a,a'}\sum_{\bk',\bq',a'',a'''}\sum_{\bar{h}}\sum_{h''}s_a s_{a''} V^{h,a,a',\bar{h}}_{\bk_1,\bk,\bk+\bq,\bk_1-\bq}V^{h'',a'',a''',h'}_{\bk'_1+\bq',\bk',\bk'+\bq',\bk'_1}\\
&\times\ll 
a^{\dagger}_{\bk}(\underline{t}_1^+) a'^{\phantom\dagger}_{\bk+\bq}(\underline{t}_1) 
a''^{\dagger}_{\bk'} (\underline{t}'^+_1) 
a'''^{\phantom\dagger}_{\bk'+\bq'}(\underline{t}'_1)
\gg_T
\ll
 \bar{h}^{\phantom\dagger}_{\bk_1-\bq}(\underline{t}_1)e^{\phantom\dagger}_{\bp_1}(\underline{t}_1)
 e'^{\dagger}_{\bp'_1}(\underline{t}'_1)h''^{\dagger}_{\bk'_1+\bq'}(\underline{t}'_1)\gg_T\\
= \frac{1}{i\hbar}&\sum_{\bk,\bq,a,a'}\sum_{\bk',\bq',a'',a'''}\sum_{\bar{h}}\sum_{h''} V^{h,a,a',\bar{h}}_{\bk_1,\bk,\bk+\bq,\bk_1-\bq}V^{h'',a'',a''',h'}_{\bk'_1+\bq',\bk',\bk'+\bq',\bk'_1}\\
&\times i\hbar
L(\bk'+\bq',-\bq',a'',a''',\underline{t}'_1,\bk+\bq,-\bq,a,a',\underline{t}_1)
i\hbar
K(1,1')\Big|_{\bk'_1+\bq',h''}^{\bk_1-\bq,\bar{h}}\\
\approx i\hbar&
\sum_{\bk,\bq,\bq',a,a'}\sum_{\bar{h},h''} V^{h,a,a',\bar{h}}_{\bk_1,\bk,\bk+\bq,\bk_1-\bq}\\
\times &\Big(
\varepsilon^{-1}(\bk'_1,-\bq',h'',h',\underline{t}'^+_1,\bk+\bq,\bq,a,a',\underline{t}_1)
-n_1\delta_{n_1,n'_1}\delta(t_1-t'_1)\delta_{\bk'_1,\bk+\bq}\delta_{\bq,-\bq'}\delta_{h'',a}\delta_{h',a'}
\Big)
K(1,1')\Big|_{\bk'_1+\bq',h''}^{\bk_1-\bq,\bar{h}}
\end{split}
\label{eq:alternative_decoupling}
\end{equation}
In the fourth line, we used the definition (\ref{eq:def_L}) of the density-density expectation value. In the fifth line, we replaced the latter
by the inverse dielectric function according to Eq.~(\ref{eq:inv_DF}), neglecting the term quadratic in the carrier density. The first term in the fifth line 
reproduces the hole-hole interaction term in the GW self-energy (\ref{eq:Sigma_GW}), while the second cancels the corresponding instantaneous (Fock-like) contribution.
We therefore note that the four-particle terms in $M(1,1')$ contain the retarded part of the excitonic GW self-energy, while any instantaneous Hartree- and Fock-type interaction are described by the $r_0$-terms. Unlike in Ref.~\cite{may_many-body_1985}, we keep the retarded part of the GW self-energy and focus on instantaneous correction terms contained in $r_0$.
\\We evaluate the $r_0$-terms in Eq.~(\ref{eq:alternative_self_energy}) with (\ref{eq:r0_explicit}) by replacing the general band summation over $a$ and $a'$ by electron and hole bands and bringing all terms to normal order. Neglecting all expectation values with four electron or hole operators, respectively, the resulting instantaneous exciton self-energy is given by:
\begin{equation}
 \begin{split}
M^{\delta}(1,1')=n_1\delta_{n_1,n'_1}\delta(t_1-t'_1)\Big( M^{\delta,(1)}(1,1')+M^{\delta,(2)}(1,1')+M^{\delta,(3)}(1,1')+M^{\delta,(4)}(1,1')  \Big)
\end{split}
\label{eq:instant_self_energy}
\end{equation}
with
\begin{equation}
 \begin{split}
M^{\delta,(1)}(1,1')=\sum_{\bk}\Big\{
&\sum_{h'',h'''}V^{h,h'',h''',h'}_{\bk_1,\bk,\bk+\bk_1-\bk'_1,\bk'_1}\delta_{\bp_1,\bp'_1}\delta_{e,e'}\ll h''^{\dagger}_{\bk}(\underline{t}_1)h'''^{\phantom\dagger}_{\bk+\bk_1-\bk'_1}(\underline{t}_1) \gg \\
-&\sum_{e'',e'''}V^{h,e'',e''',h'}_{\bk_1,\bk,\bk+\bk_1-\bk'_1,\bk'_1}\delta_{\bp_1,\bp'_1}\delta_{e,e'}\ll e''^{\dagger}_{\bk}(\underline{t}_1)e'''^{\phantom\dagger}_{\bk+\bk_1-\bk'_1}(\underline{t}_1) \gg \\
-&\sum_{h'',h'''}V^{e,h'',h''',e'}_{\bp_1,\bk,\bk+\bp_1-\bp'_1,\bp'_1}\delta_{\bk_1,\bk'_1}\delta_{h,h'}\ll h''^{\dagger}_{\bk}(\underline{t}_1)h'''^{\phantom\dagger}_{\bk+\bp_1-\bp'_1}(\underline{t}_1) \gg \\
+&\sum_{e'',e'''}V^{e,e'',e''',e'}_{\bp_1,\bk,\bk+\bp_1-\bp'_1,\bp'_1}\delta_{\bk_1,\bk'_1}\delta_{h,h'}\ll e''^{\dagger}_{\bk}(\underline{t}_1)e'''^{\phantom\dagger}_{\bk+\bp_1-\bp'_1}(\underline{t}_1) \gg
\Big\}\,,
\end{split}
\label{eq:instant_self_energy_1}
\end{equation}
\begin{equation}
 \begin{split}
M^{\delta,(2)}(1,1')=-\sum_{\bk}\Big\{
&\sum_{h'',h'''}V^{h,h'',h',h'''}_{\bk_1,\bk,\bk'_1,\bk+\bk_1-\bk'_1}\delta_{\bp_1,\bp'_1}\delta_{e,e'}\ll h''^{\dagger}_{\bk}(\underline{t}_1)h'''^{\phantom\dagger}_{\bk+\bk_1-\bk'_1}(\underline{t}_1) \gg \\
-&\sum_{e''}V^{h,e'',e',h'}_{\bk_1,\bk,\bp'_1,\bk'_1}\delta_{\bk,\bp'_1+\bk'_1-\bk_1}\ll e''^{\dagger}_{\bk}(\underline{t}_1)e^{\phantom\dagger}_{\bp_1}(\underline{t}_1) \gg \\
-&\sum_{h''}V^{e,h'',h',e'}_{\bp_1,\bk,\bk'_1,\bp'_1}\delta_{\bk,\bk'_1+\bp'_1-\bp_1}\ll h''^{\dagger}_{\bk}(\underline{t}_1)h^{\phantom\dagger}_{\bk_1}(\underline{t}_1) \gg \\
+&\sum_{e'',e'''}V^{e,e'',e',e'''}_{\bp_1,\bk,\bp'_1,\bk+\bp_1-\bp'_1}\delta_{\bk_1,\bk'_1}\delta_{h,h'}\ll e''^{\dagger}_{\bk}(\underline{t}_1)e'''^{\phantom\dagger}_{\bk+\bp_1-\bp'_1}(\underline{t}_1) \gg
\Big\}\,,
\end{split}
\label{eq:instant_self_energy_2}
\end{equation}
\begin{equation}
 \begin{split}
M^{\delta,(3)}(1,1')=\sum_{\bq}\Big\{
&\sum_{h'',h'''}V^{h,h'',h',h'''}_{\bk_1,\bk'_1-\bq,\bk'_1,\bk_1-\bq}\ll h''^{\dagger}_{\bk'_1-\bq}(\underline{t}_1)e'^{\dagger}_{\bp'_1}(\underline{t}_1) e^{\phantom\dagger}_{\bp_1}(\underline{t}_1) h'''^{\phantom\dagger}_{\bk_1-\bq}(\underline{t}_1) \gg \\
-&\sum_{e'',h'''}V^{h,e'',e',h'''}_{\bk_1,\bp'_1-\bq,\bp'_1,\bk_1-\bq}\ll h'^{\dagger}_{\bk'_1}(\underline{t}_1)e''^{\dagger}_{\bp'_1-\bq}(\underline{t}_1) e^{\phantom\dagger}_{\bp_1}(\underline{t}_1) h'''^{\phantom\dagger}_{\bk_1-\bq}(\underline{t}_1) \gg \\
-&\sum_{h'',e'''}V^{e,h'',h',e'''}_{\bp_1,\bk'_1-\bq,\bk'_1,\bp_1-\bq}\ll h''^{\dagger}_{\bk'_1-\bq}(\underline{t}_1)e'^{\dagger}_{\bp'_1}(\underline{t}_1) e'''^{\phantom\dagger}_{\bp_1-\bq}(\underline{t}_1) h^{\phantom\dagger}_{\bk_1}(\underline{t}_1) \gg \\
+&\sum_{e'',e'''}V^{e,e'',e',e'''}_{\bp_1,\bp'_1-\bq,\bp'_1,\bp_1-\bq}\ll h'^{\dagger}_{\bk'_1}(\underline{t}_1)e''^{\dagger}_{\bp'_1-\bq}(\underline{t}_1) e'''^{\phantom\dagger}_{\bp_1-\bq}(\underline{t}_1) h^{\phantom\dagger}_{\bk_1}(\underline{t}_1) \gg
\Big\}\,,
\end{split}
\label{eq:instant_self_energy_3}
\end{equation}
and
\begin{equation}
 \begin{split}
M^{\delta,(4)}(1,1')=-\sum_{\bq}\Big\{
&\sum_{h'',h'''}V^{h,h'',h''',h''}_{\bk_1,\bk'_1-\bq,\bk_1-\bq,\bk'_1}\ll h''^{\dagger}_{\bk'_1-\bq}(\underline{t}_1)e'^{\dagger}_{\bp'_1}(\underline{t}_1) e^{\phantom\dagger}_{\bp_1}(\underline{t}_1) h'''^{\phantom\dagger}_{\bk_1-\bq}(\underline{t}_1) \gg \\
-&\sum_{\bk,e'',e''',h'''}V^{h,e'',e''',h'''}_{\bk_1,\bk,\bk+\bq,\bk_1-\bq}\delta_{\bp_1,\bp'_1}\delta_{e,e'}
\ll h'^{\dagger}_{\bk'_1}(\underline{t}_1)e''^{\dagger}_{\bk}(\underline{t}_1) e'''^{\phantom\dagger}_{\bk+\bq}(\underline{t}_1) h'''^{\phantom\dagger}_{\bk_1-\bq}(\underline{t}_1) \gg \\
-&\sum_{\bk,h'',h''',e'''}V^{e,h'',h''',e'''}_{\bp_1,\bk,\bk+\bq,\bp_1-\bq}\delta_{\bk_1,\bk'_1}\delta_{h,h'}
\ll h''^{\dagger}_{\bk}(\underline{t}_1)e'^{\dagger}_{\bp'_1}(\underline{t}_1) e'''^{\phantom\dagger}_{\bp_1-\bq}(\underline{t}_1) h'''^{\phantom\dagger}_{\bk+\bq}(\underline{t}_1) \gg \\
+&\sum_{e'',e'''}V^{e,e'',e''',e'}_{\bp_1,\bp'_1-\bq,\bp_1-\bq,\bp'_1}\ll h'^{\dagger}_{\bk'_1}(\underline{t}_1)e''^{\dagger}_{\bp'_1-\bq}(\underline{t}_1) e'''^{\phantom\dagger}_{\bp_1-\bq}(\underline{t}_1) h^{\phantom\dagger}_{\bk_1}(\underline{t}_1) \gg
\Big\}\,.
\end{split}
\label{eq:instant_self_energy_4}
\end{equation}
The terms $M^{\delta,(1)}$ and $M^{\delta,(2)}$, being proportional to electron and hole occupancies, correspond to the single-particle Hartree and Fock self-energies, respectively. As discussed in detail in \cite{may_many-body_1985}, the single-particle occupancies can be approximately projected to the phase space of excitons using the relations
\begin{equation}
 \begin{split}
h^{\dagger}_{\bk}h'^{\phantom\dagger}_{\bk'}\approx \sum_{\bp,e}e^{\dagger}_{\bp}h^{\dagger}_{\bk}h'^{\phantom\dagger}_{\bk'}e^{\phantom\dagger}_{\bp} \quad
e^{\dagger}_{\bp}e'^{\phantom\dagger}_{\bp'}\approx \sum_{\bk,h}e^{\dagger}_{\bp}h^{\dagger}_{\bk}h^{\phantom\dagger}_{\bk}e'^{\phantom\dagger}_{\bp'}\,.
\end{split}
\label{eq:projection}
\end{equation}
A similar projection technique has also been used in \cite{katsch_theory_2018,erkensten_microscopic_2022}. The resulting four-particle expectation values, along with the corresponding terms in 
$M^{\delta,(3)}$ and $M^{\delta,(4)}$, can be interpreted as exciton Green functions evaluated on the time diagonal according to
\begin{equation}
 \begin{split}
i\hbar K(1,1')\Big|_{\underline{t}'_1=\underline{t}^+_1} = \ll h^{\dagger}_{\bk'_1}(\underline{t}^+_1)e'^{\dagger}_{\bp'_1}(\underline{t}^+_1) e^{\phantom\dagger}_{\bp_1}(\underline{t}_1) h^{\phantom\dagger}_{\bk_1}(\underline{t}_1) \gg\,.
\end{split}
\label{eq:exciton_time_diag}
\end{equation}

\subsubsection{Real-time self-energy in the exciton representation}

We now switch to an exciton representation of the instantaneous self-energy by introducing 
\begin{equation}
 \begin{split}
G(\nu_1,\bQ_1,\underline{t}_1,\nu'_1,\bQ'_1,\underline{t}'_1)=\sum_{\bk_1,\bp_1,\bk'_1,\bp'_1,e,h,e',h'} \Phi^*_{\nu_1,\bQ_1}(e,\bp_1,h,\bk_1) K(1,1') \Phi_{\nu'_1,\bQ'_1}(e',\bp'_1,h',\bk'_1) 
\end{split}
\label{eq:exciton_representation}
\end{equation}
with exciton wave functions as solutions to the Bethe-Salpeter equation that corresponds to the effective exciton Hamiltonian (\ref{eq:exciton_H}):
\begin{equation}
 \begin{split}
 (\varepsilon_{\bp}^e+\varepsilon_{\bk}^h  )\Phi_{\nu,\bQ}(e,\bp,h,\bk)
 -\sum_{\bq,h',e'} V^{e,h,h',e'}_{\bp,\bk,\bk+\bq,\bp-\bq} \Phi_{\nu,\bQ}(e',\bp-\bq,h',\bk+\bq)=E_{\nu,\bQ}\Phi_{\nu,\bQ}(e,\bp,h,\bk)\,.
\end{split}
\label{eq:BSE}
\end{equation}
The quantum number of the relative electron-hole motion is denoted by $\nu$, while the total exciton momentum is $\bQ=\bk+\bp$. Exciton energies are denoted by $E_{\nu,\bQ}$. Since the wave functions form a basis of the electron-hole-pair Hilbert space, the exciton representation can be inverted:
\begin{equation}
 \begin{split}
K(1,1')=\sum_{\nu_1,\nu'_1,\bQ_1,\bQ'_1} \Phi_{\nu_1,\bQ_1}(e,\bp_1,h,\bk_1) G(\nu_1,\bQ_1,\underline{t}_1,\nu'_1,\bQ'_1,\underline{t}'_1) \Phi^*_{\nu'_1,\bQ'_1}(e',\bp'_1,h',\bk'_1) \,.
\end{split}
\label{eq:exciton_representation_inv}
\end{equation}
We obtain the exciton Hartree self-energy from Eq.~(\ref{eq:instant_self_energy_1}) and the first and fourth terms of Eq.~(\ref{eq:instant_self_energy_2}) as
\begin{equation}
 \begin{split}
\Sigma^{\textrm{H}}(\nu_1,\bQ_1,\underline{t}_1,\nu'_1,\bQ'_1,\underline{t}'_1)=
i\hbar n_1\delta_{n_1,n'_1}\delta(t_1-t'_1)\sum_{\nu_2,\nu'_2,\bQ_2,\bQ'_2}\tilde{V}^{\nu_1,\nu'_2,\nu_2,\nu'_1}_{\bQ_1,\bQ'_2,\bQ_2,\bQ'_1}G(\nu_2,\bQ_2,\underline{t}_1,\nu'_2,\bQ'_2,\underline{t}_1)
\end{split}
\label{eq:exciton_Hartree_general}
\end{equation}
with the effective exciton-exciton interaction matrix elements
\begin{equation}
 \begin{split}
\tilde{V}^{\nu_1,\nu_2,\nu_3,\nu_4}_{\bQ_1,\bQ_2,\bQ_3,\bQ_4}=\tilde{V}^{\nu_1,\nu_2,\nu_3,\nu_4}_{\bQ_1,\bQ_2,\bQ=\bQ_3-\bQ_2}
=&\sum_{\bk_1,\bk_2,h_1,h_2,e_1,e_2} \Phi^*_{\nu_1,\bQ_1}(e_1,h_1,\bk_1)\Phi_{\nu_3,\bQ_2+\bQ}(e_2,h_2,\bk_2) \\
\times\Big\{
\sum_{\bar{e}}\Phi^*_{\nu_2,\bQ_2}(\bar{e},h_2,\bk_2)\Big(
&\sum_{e'} V^{e_1,\bar{e},e_2,e'}_{\bQ_1-\bk_1,\bQ_2-\bk_2,\bQ_2-\bk_2+\bQ,\bQ_1-\bk_1-\bQ} \Phi_{\nu_4,\bQ_1-\bQ}(e',h_1,\bk_1) 
 \\
-&\sum_{e'} V^{e_1,\bar{e},e',e_2}_{\bQ_1-\bk_1,\bQ_2-\bk_2,\bQ_1-\bk_1-\bQ,\bQ_2-\bk_2+\bQ} \Phi_{\nu_4,\bQ_1-\bQ}(e',h_1,\bk_1) 
 \\
-&\sum_{h'} V^{h_1,\bar{e},e_2,h'}_{\bk_1,\bQ_2-\bk_2,\bQ_2-\bk_2+\bQ,\bk_1-\bQ} \Phi_{\nu_4,\bQ_1-\bQ}(e_1,h',\bk_1-\bQ) 
\Big) \\
+\sum_{\bar{h}}\Phi^*_{\nu_2,\bQ_2}(e_2,\bar{h},\bk_2-\bQ)\Big(
&\sum_{h'} V^{h_1,\bar{h},h_2,h'}_{\bk_1,\bk_2-\bQ,\bk_2,\bk_1-\bQ} \Phi_{\nu_4,\bQ_1-\bQ}(e_1,h',\bk_1-\bQ) 
 \\
-&\sum_{h'} V^{h_1,\bar{h},h',h_2}_{\bk_1,\bk_2-\bQ,\bk_1-\bQ,\bk_2} \Phi_{\nu_4,\bQ_1-\bQ}(e_1,h',\bk_1-\bQ) 
 \\
-&\sum_{e'} V^{e_1,\bar{h},h_2,e'}_{\bQ_1-\bk_1,\bk_2-\bQ,\bk_2,\bQ_1-\bk_1-\bQ} \Phi_{\nu_4,\bQ_1-\bQ}(e',h_1,\bk_1) 
\Big)
\Big\}\\
=&\tilde{V}^{\textrm{(D)},\nu_1,\nu_2,\nu_3,\nu_4}_{\bQ_1,\bQ_2,\bQ=\bQ_3-\bQ_2}-\tilde{V}^{\textrm{(X)},\nu_1,\nu_2,\nu_3,\nu_4}_{\bQ_1,\bQ_2,\bQ=\bQ_3-\bQ_2}\,.
\end{split} 
\label{eq:V_eff}
\end{equation}
Note that we have made use of the momentum conservation relation $\bp=\bQ-\bk$ implied in the exciton wave functions to eliminate electron momenta. Terms number 1, 3, 4, and 6 stem from the single-particle Hartree term and correspond to the semiclassical dipolar Coulomb interaction between excitons \cite{schindler_analysis_2008}, which is described by $\tilde{V}^{\textrm{(D)}}$. We find two repulsive and two attractive contributions caused by the opposite charge of electrons and holes. Terms number 2 and 5, however, are non-classical terms due to the fermionic nature of the exciton constituents: These fermionic correction terms describe the exchange of either an electron or a hole between two excitons leading to an attractive interaction proportional to $\tilde{V}^{\textrm{(X)}}$.
Notably, besides the sign the only difference between ``direct'' and ``exchange'' terms inside the effective matrix element is the swap of two states in the elementary Coulomb matrix element.
\\In a similar way, an exciton Fock self-energy is obtained from Eq.~(\ref{eq:instant_self_energy_3}) as well as the first and fourth terms of Eq.~(\ref{eq:instant_self_energy_4}):
\begin{equation}
 \begin{split}
\Sigma^{\textrm{F}}(\nu_1,\bQ_1,\underline{t}_1,\nu'_1,\bQ'_1,\underline{t}'_1)=
i\hbar n_1\delta_{n_1,n'_1}\delta(t_1-t'_1)\sum_{\nu_2,\nu'_2,\bQ_2,\bQ'_2}\tilde{V}^{\nu_1,\nu'_2,\nu'_1,\nu_2}_{\bQ_1,\bQ'_2,\bQ'_1,\bQ_2}G(\nu_2,\bQ_2,\underline{t}_1,\nu'_2,\bQ'_2,\underline{t}_1)\,.
\end{split}
\label{eq:exciton_Fock_general}
\end{equation}
The exciton Fock self-energy can be interpreted as resulting from the bosonic exchange of two excitons, with only two states in the effective matrix element swapped with respect to the exciton Hartree term. As the latter, the Fock term contains fermionic exchange terms.
\\The remaining four terms of the instantaneous self-energy, which are the second and third terms of (\ref{eq:instant_self_energy_2}) and (\ref{eq:instant_self_energy_4}), respectively, can be grouped together similar to the exciton Hartree and Fock self-energies:
\begin{equation}
 \begin{split}
\Sigma^{\textrm{PB}}(\nu_1,\bQ_1,\underline{t}_1,\nu'_1,\bQ'_1,\underline{t}'_1)=
i\hbar n_1\delta_{n_1,n'_1}\delta(t_1-t'_1)\sum_{\nu_2,\nu'_2,\bQ_2,\bQ'_2}\Big(
\tilde{V}^{\textrm{PB},\nu_1,\nu'_2,\nu_2,\nu'_1}_{\bQ_1,\bQ'_2,\bQ_2,\bQ'_1}
+\tilde{V}^{\textrm{PB},\nu_1,\nu'_2,\nu'_1,\nu_2}_{\bQ_1,\bQ'_2,\bQ'_1,\bQ_2}
\Big)G(\nu_2,\bQ_2,\underline{t}_1,\nu'_2,\bQ'_2,\underline{t}_1)\,.
\end{split}
\label{eq:exciton_PB_general}
\end{equation}
We introduced new effective matrix elements
\begin{equation}
 \begin{split}
\tilde{V}^{\textrm{PB},\nu_1,\nu_2,\nu_3,\nu_4}_{\bQ_1,\bQ_2,\bQ_3,\bQ_4}=&\tilde{V}^{\textrm{PB},\nu_1,\nu_2,\nu_3,\nu_4}_{\bQ_1,\bQ_2,\bar{\bQ}=\bQ_4-\bQ_2}
=\sum_{\bk_1,\bk_2,h_1,h_2,e_1,e_2} \Phi^*_{\nu_1,\bQ_1}(e_1,h_1,\bk_1)\\
\times\Big\{&\Phi_{\nu_4,\bQ_2+\bar{\bQ}}(e_2,h_2,\bk_2) 
\sum_{\bar{h},e''} V^{h_1,e'',e_2,h_2}_{\bk_1,\bQ_2+\bar{\bQ}-\bk_1,\bQ_2+\bar{\bQ}-\bk_2,\bk_2} \Phi_{\nu_3,\bQ_1-\bar{\bQ}}(e_1,\bar{h},\bk_1-\bar{\bQ})  \Phi^*_{\nu_2,\bQ_2}(e'',\bar{h},\bk_1-\bar{\bQ})
 \\
+&\Phi_{\nu_3,\bQ_1-\bar{\bQ}}(e_2,h_2,\bk_2-\bar{\bQ}) 
\sum_{\bar{e},h''} V^{h'',e_1,e_2,h_2}_{\bk_1-\bar{\bQ},\bQ_1-\bk_1,\bQ_1-\bk_2,\bk_2-\bar{\bQ}} \Phi_{\nu_4,\bQ_2+\bar{\bQ}}(\bar{e},h_1,\bk_1)  \Phi^*_{\nu_2,\bQ_2}(\bar{e},h'',\bk_1-\bar{\bQ})
\Big\}\,.
\end{split} 
\label{eq:V_eff_PB}
\end{equation}
The self-energy contribution (\ref{eq:exciton_PB_general}) appears on the same level as the Hartree-Fock energy renormalizations, is proportional to electron-hole interaction matrix elements and has the positive sign of a repulsive interaction. It is therefore plausible to attribute the self-energy to Pauli blocking caused by phase-space filling from the fermionic constituents. In a Bethe-Salpeter or Wannier equation for the exciton, phase-space filling appears as a prefactor similar to (\ref{eq:PSF_factor}) of the electron-hole term, thereby reducing the exciton binding energy \cite{steinhoff_influence_2014}. Expanding the density-like terms in the phase-space filling factor according to Eq.~(\ref{eq:projection}) and adding the corresponding bosonic exchange term, $\Sigma^{\textrm{PB}}$ can be derived. We note that unlike the fermionic exchange corrections in $\Sigma^{\textrm{H}}$ and $\Sigma^{\textrm{F}}$, $\Sigma^{\textrm{PB}}$ can not simply be obtained by swapping two states in a Coulomb matrix element. Both, the fermionic correction terms and the Pauli-blocking terms are beyond the RPA self-energy derived in the previous subsection via functional derivative technique. 
\\We now bring the retarded part of the exciton GW self-energy to a form similar to the above exciton Hartree and Fock self-energies. Starting from Eq.~(\ref{eq:alternative_decoupling}), we derive an expression that contains the fully screened Coulomb interaction $W$. Based on the arrangement of operators in the definition (\ref{eq:inv_DF}), the elements of the dielectric matrix can be identified as
\begin{equation}
 \begin{split}
\varepsilon^{-1}(\bk_2+\bq_2,\bq_2,a_2,a'_2,\underline{t}_2,\bk+\bq,\bq,a,a',\underline{t}_1)=& 
\big<\psi_{\bk_2}^{a_2}\big|\big<\psi_{\bk+\bq}^{a'}\big| \varepsilon^{-1}(\br,\br',\underline{t}_2,\underline{t}_1) \big|\psi_{\bk}^{a}\big>\big|\psi_{\bk_2+\bq_2}^{a'_2}\big> \\
\approx&\varepsilon^{-1,a_2,a',a,a'_2}_{\bk_2,\bk+\bq,\bk,\bk_2+\bq_2}(\underline{t}_2,\underline{t}_1)\delta_{\bq,\bq_2}\delta_{\bk,\bk_2}
\,.
\end{split}
\label{eq:inv_DF_elements}
\end{equation}
in analogy to Coulomb matrix elements. The first Kronecker delta takes momentum conservation into account, while the second is an additional approximation. Expanding the dielectric and Coulomb matrix elements in terms of Wannier functions and making use of the completeness of Bloch states, we find
\begin{equation}
 \begin{split}
&i\hbar
\sum_{\bk,\bq,\bq',a,a'}\sum_{\bar{h},h''} V^{h,a,a',\bar{h}}_{\bk_1,\bk,\bk+\bq,\bk_1-\bq}
\varepsilon^{-1}(\bk'_1,-\bq',h'',h',\underline{t}'^+_1,\bk+\bq,\bq,a,a',\underline{t}_1)
K(1,1')\Big|_{\bk'_1+\bq',h''}^{\bk_1-\bq,\bar{h}} \\
=&i\hbar
\sum_{\bq,a,a'}\sum_{\bar{h},h''} V^{h,a,a',\bar{h}}_{\bk_1,\bk'_1-\bq,\bk'_1,\bk_1-\bq}
\varepsilon^{-1,h'',a',a,h'}_{\bk'_1-\bq,\bk'_1,\bk'_1-\bq,\bk'_1}(\underline{t}'^+_1,\underline{t}_1)
K(1,1')\Big|_{\bk'_1-\bq,h''}^{\bk_1-\bq,\bar{h}} \\
=&i\hbar
\sum_{\bq}\sum_{\bar{h},h''} \sum_{\alpha,\alpha'} 
\big( c_{\alpha', \bk'_1-\bq}^{h''} \big)^{*} \big( c_{\alpha, \bk_1}^{h} \big)^{*} W^{\alpha'\alpha}_{-\bq}(\underline{t}'^+_1,\underline{t}_1)c_{\alpha, \bk_1-\bq}^{\bar{h}} c_{\alpha', \bk'_1}^{h'}
K(1,1')\Big|_{\bk'_1-\bq,h''}^{\bk_1-\bq,\bar{h}} \\
=&i\hbar \sum_{\bq}\sum_{\bar{h},h''} 
W^{h'',h,\bar{h},h'}_{\bk'_1-\bq,\bk_1,\bk_1-\bq,\bk'_1}(\underline{t}'^+_1,\underline{t}_1)
K(1,1')\Big|_{\bk'_1-\bq,h''}^{\bk_1-\bq,\bar{h}} 
\end{split}
\label{eq:GW_Wannier}
\end{equation}
with the fully screened Coulomb matrix 
\begin{equation}
\begin{split} 
  W^{\alpha\beta}_{\bq}(\underline{t},\underline{t}')=\sum_{\gamma} \varepsilon^{-1,\alpha\gamma}_{\bq}(\underline{t},\underline{t}')V^{\gamma\beta}_{\bq}
     \,.
    \label{eq:W_matrix}
\end{split}
\end{equation}
Analogous calculations are carried out for the three remaining four-particle terms in Eq.~(\ref{eq:alternative_self_energy}). Then we arrive at the exciton GW self-energy:
\begin{equation}
 \begin{split}
\Sigma^{\textrm{GW}}(\nu_1,\bQ_1,\underline{t}_1,\nu'_1,\bQ'_1,\underline{t}'_1)=
i\hbar\sum_{\nu_2,\nu'_2,\bQ_2,\bQ'_2}\tilde{W}^{\nu_1,\nu'_2,\nu'_1,\nu_2}_{\bQ_1,\bQ'_2,\bQ'_1,\bQ_2}(\underline{t}'_1,\underline{t}_1)G(\nu_2,\bQ_2,\underline{t}_1,\nu'_2,\bQ'_2,\underline{t}'_1)
\end{split}
\label{eq:exciton_GW_general}
\end{equation}
with the effective matrix element
\begin{equation}
 \begin{split}
\tilde{W}^{\nu_1,\nu_2,\nu_3,\nu_4}_{\bQ_1,\bQ_2,\bQ=\bQ_3-\bQ_2}(\underline{t}'_1,\underline{t}_1)
=&\sum_{\bk_1,\bk_2,h_1,h_2,e_1,e_2} \Phi^*_{\nu_1,\bQ_1}(e_1,h_1,\bk_1)\Phi_{\nu_3,\bQ_2+\bQ}(e_2,h_2,\bk_2) \\
\times\Big\{
\sum_{\bar{e}}\Phi^*_{\nu_2,\bQ_2}(\bar{e},h_2,\bk_2)\Big(
&\sum_{e'} W^{\bar{e},e_1,e',e_2}_{\bQ_2-\bk_2,\bQ_1-\bk_1,\bQ_1-\bk_1-\bQ,\bQ_2-\bk_2+\bQ}(\underline{t}'_1,\underline{t}_1) \Phi_{\nu_4,\bQ_1-\bQ}(e',h_1,\bk_1) 
 \\
-&\sum_{h'} W^{\bar{e},h_1,h',e_2}_{\bQ_2-\bk_2,\bk_1,\bk_1-\bQ,\bQ_2-\bk_2+\bQ}(\underline{t}'_1,\underline{t}_1) \Phi_{\nu_4,\bQ_1-\bQ}(e_1,h',\bk_1-\bQ) 
\Big) \\
+\sum_{\bar{h}}\Phi^*_{\nu_2,\bQ_2}(e_2,\bar{h},\bk_2-\bQ)\Big(
&\sum_{h'} W^{\bar{h},h_1,h',h_2}_{\bk_2-\bQ,\bk_1,\bk_1-\bQ,\bk_2}(\underline{t}'_1,\underline{t}_1) \Phi_{\nu_4,\bQ_1-\bQ}(e_1,h',\bk_1-\bQ) 
 \\
-&\sum_{e'} W^{\bar{h},e_1,e',h_2}_{\bk_2-\bQ_2,\bQ_1-\bk_1,\bQ_1-\bk_1-\bQ,\bk_2}(\underline{t}'_1,\underline{t}_1) \Phi_{\nu_4,\bQ_1-\bQ}(e',h_1,\bk_1) 
\Big)
\Big\}\,.
\end{split} 
\label{eq:W_eff}
\end{equation}
\\To summarize, we have obtained a Dyson-type equation for the exciton Green function by upgrading Eq.~(\ref{eq:exciton_GF_full_EOM_integral_alternative}):
\begin{equation}
 \begin{split}
\int_{\mathcal{C}}d\underline{t}_2 \sum_{\bk_2,\bp_2,h_2,e_2 }K^{-1}_0(1,2) K(2,1') = n_1\delta_{n_1,n'_1}\delta(t_1-t'_1)F(1,1')
+\int_{\mathcal{C}}d\underline{t}_2 \sum_{\bk_2,\bp_2,h_2,e_2 } M(1,2) K(2,1')\, .
\end{split}
\label{eq:exciton_Dyson}
\end{equation}
The exciton self-energy $M(1,2)$ contains Hartree- and Fock-type contributions including fermionic corrections, a Pauli-blocking term as well as an RPA-type correlation term stemming from the factorization of four-particle Green functions. 
Consistent with the derivation of the RPA vertex function in Eqs.~(\ref{eq:inverse_exciton_GF_Dyson}) and (\ref{eq:vertexfct_approx}), we drop the density-like terms in $F(1,1')$. 
Then, using the Bethe-Salpeter equation (\ref{eq:BSE}), Dyson's equation can be expressed in the exciton representation:
\begin{equation}
 \begin{split}
&\int_{\mathcal{C}}d\underline{t}_2 \sum_{\nu_2,\bQ_2} \Big\{  \Big(i\hbar\frac{\partial}{\partial t_1} - E_{\nu_1,\bQ_1}\Big)\delta_{\nu_1,\nu_2}\delta_{\bQ_1,\bQ_2} n_1\delta_{n_1,n_2}\delta(t_1-t_2) \\ - 
&\Sigma^{\textrm{H}}(\nu_1,\bQ_1,\underline{t}_1,\nu_2,\bQ_2,\underline{t}_2)
- \Sigma^{\textrm{F}}(\nu_1,\bQ_1,\underline{t}_1,\nu_2,\bQ_2,\underline{t}_2)
- \Sigma^{\textrm{PB}}(\nu_1,\bQ_1,\underline{t}_1,\nu_2,\bQ_2,\underline{t}_2)
\Big\} G(\nu_2,\bQ_2,\underline{t}_2,\nu'_1,\bQ'_1,\underline{t}'_1) \\ = & n_1\delta_{n_1,n'_1}\delta(t_1-t'_1)\delta_{\nu_1,\nu'_1}\delta_{\bQ_1,\bQ'_1} \\
+&\int_{\mathcal{C}}d\underline{t}_2 \sum_{\nu_2,\bQ_2}
\Big(
\Sigma^{\textrm{GW}}(\nu_1,\bQ_1,\underline{t}_1,\nu_2,\bQ_2,\underline{t}_2)
-\Sigma^{\textrm{GW},\delta}(\nu_1,\bQ_1,\underline{t}_1,\nu_2,\bQ_2,\underline{t}_2)
\Big)
G(\nu_2,\bQ_2,\underline{t}_2,\nu'_1,\bQ'_1,\underline{t}'_1)\, .
\end{split}
\label{eq:exciton_Dyson_X_picture}
\end{equation}
Here $\Sigma^{\textrm{GW},\delta}$ denotes the instantaneous part of the exciton-GW self-energy emerging from the factorization of four-particle Green functions according to Eq.~(\ref{eq:alternative_decoupling}). As described in Refs.~\cite{may_many-body_1985,boldt_many-body_1985,kremp_quantum_2005}, we transform Dyson's equation to physical times by unfolding the Keldysh contour $(\int_{\mathcal{C}}d\underline{t}_2=\sum_{n_2} \int_{-\infty}^{\infty}dt_2)$, dropping the external potential and using the Langreth-Wilkins theorems. In close analogy to single-particle Green functions, we introduce greater and lesser Green functions
\begin{equation}
 \begin{split}
g^>(\nu_1,\bQ_1,t_1,\nu'_1,\bQ'_1,t'_1) & = G^{n_1=-,n'_1=+}(\nu_1,\bQ_1,t_1,\nu'_1,\bQ'_1,t'_1) \\ 
g^<(\nu_1,\bQ_1,t_1,\nu'_1,\bQ'_1,t'_1) & = G^{n_1=+,n'_1=-}(\nu_1,\bQ_1,t_1,\nu'_1,\bQ'_1,t'_1)\,
\end{split}
\label{eq:greater_lesser}
\end{equation}
and retarded Green functions 
\begin{equation}
 \begin{split}
g^{\textrm{ret}}(\nu_1,\bQ_1,t_1,\nu'_1,\bQ'_1,t'_1)  = 
\theta(t_1-t'_1)\Big(
g^>(\nu_1,\bQ_1,t_1,\nu'_1,\bQ'_1,t'_1) 
-g^<(\nu_1,\bQ_1,t_1,\nu'_1,\bQ'_1,t'_1)
\Big)
\, .
\end{split}
\label{eq:ret_G}
\end{equation}
Retarded self-energies additionally contain an instantaneous contribution:
\begin{equation}
 \begin{split}
\Sigma^{\textrm{ret}}(\nu_1,\bQ_1,t_1,\nu'_1,\bQ'_1,t'_1)  = 
\Sigma^{\delta}(\nu_1,\bQ_1,t_1,\nu'_1,\bQ'_1,t'_1)+\theta(t_1-t'_1)\Big(
\Sigma^>(\nu_1,\bQ_1,t_1,\nu'_1,\bQ'_1,t'_1) 
-\Sigma^<(\nu_1,\bQ_1,t_1,\nu'_1,\bQ'_1,t'_1)
\Big)
\, .
\end{split}
\label{eq:ret_Sigma}
\end{equation}
In the following, we assume that Green functions and self-energies are diagonal in the exciton representation. Equal-time Green functions, which appear in the instantaneous self-energy, correspond to $g^<$-functions, consistent with the replacement (\ref{eq:exciton_time_diag}). We further assume that the exciton gas is in a quasi-equilibrium state, where $g^<$ is a time-independent distribution function. We again discard Pauli-blocking nonlinearities by describing excitons as ideal bosons, which implies
\begin{equation}
 \begin{split}
i\hbar g^<(\nu_1,\bQ_1)=N^{\textrm{X}}_{\nu_1,\bQ_1}=\Big[\textrm{exp}\Big(\frac{E_{\nu_1,\bQ_1} - \mu_{\textrm{X}}}{k_{\textrm{B}} T}\Big)-1\Big]^{-1}
\end{split}
\label{eq:N_X}
\end{equation}
with the temperature $T$ and the exciton chemical potential $\mu_{\textrm{X}} $. According to the bosonic commutator rules, it is
\begin{equation}
 \begin{split}
i\hbar g^>(\nu_1,\bQ_1)=1+i\hbar g^<(\nu_1,\bQ_1)\,.
\end{split}
\label{eq:N_X_plus_1}
\end{equation}
In the quasi-equilibrium case, evaluating $i\hbar\frac{\partial}{\partial t_1} g^{\textrm{ret}}(\nu_1,\bQ_1,t_1)$ with Eqs.~(\ref{eq:ret_G}) and (\ref{eq:exciton_Dyson_X_picture}), an algebraic equation for the retarded exciton Green function can be derived in the frequency domain:
\begin{equation}
 \begin{split}
\Big( \hbar\omega - E_{\nu_1,\bQ_1} - 
\Sigma^{\textrm{H}}(\nu_1,\bQ_1)
- \Sigma^{\textrm{F}}(\nu_1,\bQ_1)
- \Sigma^{\textrm{PB}}(\nu_1,\bQ_1)
- \Sigma^{\textrm{MW,ret}}(\nu_1,\bQ_1,\omega)
\Big) g^{\textrm{ret}}(\nu_1,\bQ_1,\omega)  = 1
\, .
\end{split}
\label{eq:ret_G_freq}
\end{equation}
The retarded Green function describes the spectral properties of the dense exciton gas. Many-body interaction effects are taken into account via the self-energy $\Sigma(\omega)=\Sigma^{\textrm{H}}
+ \Sigma^{\textrm{F}}
+ \Sigma^{\textrm{PB}}
+ \Sigma^{\textrm{MW,ret}}(\omega) $, which acts as a frequency-dependent operator. We have introduced a Montroll-Ward self-energy $\Sigma^{\textrm{MW,ret}}(\omega)$ for excitons, which contains only the retarded part of the GW self-energy (\ref{eq:exciton_GW_general}). Assuming that excitons can be described as quasi-particles with a 
renormalized energy $\tilde{E}_{\nu_1,\bQ_1}$ and a broadening $\Gamma_{\nu_1,\bQ_1}$, the exciton Montroll-Ward self-energy can be derived along the same lines as in the single-particle case \cite{kremp_quantum_2005,steinhoff_exciton_2017}:
\begin{equation}
 \begin{split}
\Sigma^{\textrm{MW,ret}}(\nu_1,\bQ_1,\omega)=i\hbar\sum_{\nu'_1,\bQ'_1}\int_{-\infty}^{\infty}\frac{d\omega'}{2\pi}\frac{(1+N^{\textrm{X}}_{\nu'_1,\bQ'_1})\tilde{W}^{>,\nu_1,\nu'_1,\nu_1,\nu'_1}_{\bQ_1,\bQ'_1,\bQ_1,\bQ'_1}(\omega')-N^{\textrm{X}}_{\nu'_1,\bQ'_1}\tilde{W}^{<,\nu_1,\nu'_1,\nu_1,\nu'_1}_{\bQ_1,\bQ'_1,\bQ_1,\bQ'_1}(\omega')}{\hbar\omega-\tilde{E}_{\nu'_1,\bQ'_1}+i \Gamma_{\nu'_1,\bQ'_1}-\hbar\omega'}
\,.
\end{split}
\label{eq:MW_exciton}
\end{equation}
The plasmon propagators $\tilde{W}^{>/<}(\omega)$ are given by the effective fully screened matrix elements (\ref{eq:W_eff}), which are linear combinations of Coulomb matrix elements $W$ in the Bloch basis. Using the unitary transformation to Wannier orbitals as shown in Eq.~(\ref{eq:GW_Wannier}), the exciton Montroll-Ward self-energy can be expressed in terms of plasmon propagators in the Wannier basis.
The latter fulfill the Kubo-Martin-Schwinger relation \cite{kremp_quantum_2005}
\begin{equation}
\begin{split} 
  W^{>}_{\alpha\beta,\bq}(\omega)= e^{\frac{\hbar\omega}{k_{\textrm{B}} T}} W^{<}_{\alpha\beta,\bq}(\omega)\,.
     \,,
    \label{eq:W_prop_KMS}
\end{split}
\end{equation}
which in combination with $2i\textrm{Im}\,W^{\textrm{ret}}_{\alpha\beta,\bq}(\omega)=W^{>}_{\alpha\beta,\bq}(\omega)-W^{<}_{\alpha\beta,\bq}(\omega) $ allows to directly relate the propagators to the retarded Coulomb interaction:
\begin{equation}
\begin{split} 
  W^{>}_{\alpha\beta,\bq}(\omega)&=(1+n_{\textrm{B}}(\omega))\,2i\,\textrm{Im}\,W^{\textrm{ret}}_{\alpha\beta,\bq}(\omega), \\
  W^{<}_{\alpha\beta,\bq}(\omega)&=   n_{\textrm{B}}(\omega) \,2i\,\textrm{Im}\,W^{\textrm{ret}}_{\alpha\beta,\bq}(\omega)
     \,
    \label{eq:W_prop_ret}
\end{split}
\end{equation}
with the Bose distribution function $n_{\textrm{B}}(\omega)$.
The retarded fully screened Coulomb matrix is obtained using the inverse dielectric matrix for photo-excited carriers according to Eq.~(\ref{eq:W_matrix}):
\begin{equation}
\begin{split} 
  W^{\textrm{ret}}_{\alpha\beta,\bq}(\omega)&=\sum_{\gamma} \varepsilon^{-1,\textrm{ret},\alpha\gamma}_{\bq}(\omega)V^{\gamma\beta}_{\bq}
     \,.
    \label{eq:W_ret_eps}
\end{split}
\end{equation}
The dielectric matrix itself is given by
\begin{equation}
\begin{split} 
  \varepsilon^{\textrm{ret},\alpha\beta}_{\bq}(\omega)=\delta_{\alpha,\beta}-\sum_{\gamma}V^{\alpha\gamma}_{\bq} \Pi^{\textrm{ret},\gamma\beta}_{\bq}(\omega)
    \label{eq:eps_ab}
\end{split}
\end{equation}
and will be discussed in the following subsection.
\\ The instantaneous part of the self-energy is derived from Eqs.~(\ref{eq:exciton_Hartree_general}), (\ref{eq:exciton_Fock_general}) and (\ref{eq:exciton_PB_general}) using the approximation for the equal-time (lesser) Green functions (\ref{eq:N_X}) and the splitting of effective matrix elements (\ref{eq:V_eff}) into direct and exchange parts:
\begin{subequations}
\begin{align}
\Sigma^{\textrm{H}}(\nu_1,\bQ_1)&=
\sum_{\nu'_1,\bQ'_1}\tilde{V}^{\nu_1,\nu'_1,\nu'_1,\nu_1}_{\bQ_1,\bQ'_1,\bQ'_1,\bQ_1}N^{\textrm{X}}_{\nu'_1,\bQ'_1}
=\Sigma^{\textrm{H,(D)}}(\nu_1,\bQ_1)+\Sigma^{\textrm{H,(X)}}(\nu_1,\bQ_1)\, , \label{eq:exciton_Hartree_final} \\
\Sigma^{\textrm{H,(D)}}(\nu_1,\bQ_1)&=
\sum_{\nu'_1,\bQ'_1}\tilde{V}^{\textrm{(D)},\nu_1,\nu'_1,\nu'_1,\nu_1}_{\bQ_1,\bQ'_1,\bQ'_1,\bQ_1}N^{\textrm{X}}_{\nu'_1,\bQ'_1}\, , \label{eq:exciton_Hartree_final_direct} \\
\Sigma^{\textrm{H,(X)}}(\nu_1,\bQ_1)&=
-\sum_{\nu'_1,\bQ'_1}\tilde{V}^{\textrm{(X)},\nu_1,\nu'_1,\nu'_1,\nu_1}_{\bQ_1,\bQ'_1,\bQ'_1,\bQ_1}N^{\textrm{X}}_{\nu'_1,\bQ'_1}\, , \label{eq:exciton_Hartree_final_X}
\end{align}
\end{subequations}
\begin{subequations}
\begin{align}
\Sigma^{\textrm{F}}(\nu_1,\bQ_1)&=
\sum_{\nu'_1,\bQ'_1}\tilde{V}^{\nu_1,\nu'_1,\nu_1,\nu'_1}_{\bQ_1,\bQ'_1,\bQ_1,\bQ'_1}N^{\textrm{X}}_{\nu'_1,\bQ'_1}
=\Sigma^{\textrm{F,(D)}}(\nu_1,\bQ_1)+\Sigma^{\textrm{F,(X)}}(\nu_1,\bQ_1)\, , \label{eq:exciton_Fock_final} \\
\Sigma^{\textrm{F,(D)}}(\nu_1,\bQ_1)&=
\sum_{\nu'_1,\bQ'_1}\tilde{V}^{\textrm{(D)},\nu_1,\nu'_1,\nu_1,\nu'_1}_{\bQ_1,\bQ'_1,\bQ_1,\bQ'_1}N^{\textrm{X}}_{\nu'_1,\bQ'_1}\, , \label{eq:exciton_Fock_final_direct} \\
\Sigma^{\textrm{F,(X)}}(\nu_1,\bQ_1)&=
-\sum_{\nu'_1,\bQ'_1}\tilde{V}^{\textrm{(X)},\nu_1,\nu'_1,\nu_1,\nu'_1}_{\bQ_1,\bQ'_1,\bQ_1,\bQ'_1}N^{\textrm{X}}_{\nu'_1,\bQ'_1}\, , \label{eq:exciton_Fock_final_X}
\end{align}
\end{subequations}
\begin{equation}
 \begin{split}
\Sigma^{\textrm{PB}}(\nu_1,\bQ_1)=
\sum_{\nu'_1,\bQ'_1}
\Big(
\tilde{V}^{\textrm{PB},\nu_1,\nu'_1,\nu'_1,\nu_1}_{\bQ_1,\bQ'_1,\bQ'_1,\bQ_1}
+\tilde{V}^{\textrm{PB},\nu_1,\nu'_1,\nu_1,\nu'_1}_{\bQ_1,\bQ'_1,\bQ_1,\bQ'_1}
\Big)
N^{\textrm{X}}_{\nu'_1,\bQ'_1}\, .
\end{split}
\label{eq:exciton_PB_final}
\end{equation}
Due to the Coulomb singularity at long wavelength ($\bQ=0$ in Eq.~(\ref{eq:V_eff})), the Hartree interaction requires a separate treatment in the Wannier representation. 
For a charge-neutral system the macroscopic (leading) term drops out and only microscopic contributions to Hartree interaction remain. Following the procedure in \cite{schobert_ab_2023} we calculate Hartree-type matrix elements by setting the macroscopic eigenvalue of the Coulomb matrix to zero before transforming the matrix element to the Wannier representation. This means that the dipole-dipole interaction is not screened by the dielectric environment but only by the polarizability of the bilayer itself. This observation is consistent with the model developed in Ref.~\cite{erkensten_microscopic_2022}. 
\\The quasi-particle approximation for exciton Green functions implies that the exciton self-energy $\Sigma(\nu_1,\bQ_1,\omega)=\Sigma^{\textrm{H}}(\nu_1,\bQ_1)
+ \Sigma^{\textrm{F}}(\nu_1,\bQ_1)
+ \Sigma^{\textrm{PB}}(\nu_1,\bQ_1)
+ \Sigma^{\textrm{MW,ret}}(\nu_1,\bQ_1,\omega) $
has to be evaluated self-consistently:
\begin{equation}
 \begin{split}
\tilde{E}_{\nu_1,\bQ_1} & = E_{\nu_1,\bQ_1} + \textrm{Re}\, \Sigma(\nu_1,\bQ_1,\tilde{E}_{\nu_1,\bQ_1} / \hbar)\, , \\
\Gamma_{\nu_1,\bQ_1}    & = - \textrm{Im}\, \Sigma(\nu_1,\bQ_1,\tilde{E}_{\nu_1,\bQ_1} / \hbar) + \Gamma_0 \, .
\end{split}
\label{eq:Sigma_self_cons}
\end{equation}
Since the instantaneous part of the self-energy is real-valued, quasi-particle broadening induced by exciton-exciton interaction stems only from the Montroll-Ward self-energy. We additionally introduce a phenomenological broadening $\Gamma_0$ that takes into account scattering of excitons with phonons and defects. We choose $\Gamma_0=5$ meV independent of the temperature.
\\ Finally, we note that a static limit can be applied to the self-energy similar to the single-particle case \cite{erben_excitation-induced_2018}, which results in a screened-exchange-Coulomb-hole (SXCH) self-energy for excitons:
\begin{equation}
 \begin{split}
&\Sigma^{\textrm{F,(D)}}(\nu_1,\bQ_1) + \Sigma^{\textrm{MW,ret}}(\nu_1,\bQ_1,\tilde{E}_{\nu_1,\bQ_1} / \hbar)  \approx \Sigma^{\textrm{SXCH}}(\nu_1,\bQ_1) \\ 
& = \sum_{\nu'_1,\bQ'_1}\tilde{W}^{\nu_1,\nu'_1,\nu_1,\nu'_1}_{\bQ_1,\bQ'_1,\bQ_1,\bQ'_1}(\omega=0)N^{\textrm{X}}_{\nu'_1,\bQ'_1}
+\frac{1}{2}\sum_{\nu'_1,\bQ'_1}\Big(\tilde{W}^{\nu_1,\nu'_1,\nu_1,\nu'_1}_{\bQ_1,\bQ'_1,\bQ_1,\bQ'_1}(\omega=0)-\tilde{V}^{\textrm{(D)},\nu_1,\nu'_1,\nu_1,\nu'_1}_{\bQ_1,\bQ'_1,\bQ_1,\bQ'_1}\Big)\,.
\end{split}
\label{eq:Sigma_SXCH}
\end{equation}
From the structure of the exciton Montroll-Ward self-energy, it follows that (static) screening due to photo-excited carriers is applied to the bosonic exchange interaction described by the exciton Fock self-energy. However, 
the fermionic correction terms to the Fock self-energy are not screened. Also, the exciton Hartree and Pauli-blocking contributions remain unscreened. Besides screening of excitonic exchange, a Coulomb hole self-energy arises and leads to a red shift of energies. Throughout this work, we take into account the full frequency dependence of screening, which is discussed in detail in the following.

\subsubsection{Excitonic screening in RPA}

Consistent with the inverse dielectric function (\ref{eq:inv_DF}), the dielectric function itself is given by
\begin{equation}
 \begin{split}
&\varepsilon(\bk+\bq,\bq,a,a',\underline{t}_1,\bk_2+\bq_2,\bq_2,a_2,a'_2,\underline{t}_2)=\frac{\delta V^{a, a'}_{\textrm{ext}}(\bk+\bq,\bq,\underline{t}_1)}{\delta V^{a_2, a'_2}_{\textrm{eff}}(\bk_2+\bq_2,\bq_2,\underline{t}_2)}
\end{split}
\label{eq:DF}
\end{equation}
with the relation between $V_{\textrm{ext}} $ and $V_{\textrm{eff}} $ given in Eq.~(\ref{eq:V_eff_pot}). Using the definition of the partice-density operator (\ref{eq:dens_op}), we obtain:
\begin{equation}
 \begin{split}
&\varepsilon(\bk+\bq,\bq,a,a',\underline{t}_1,\bk_2+\bq_2,\bq_2,a_2,a'_2,\underline{t}_2)\\
=&n_1\delta_{n_1,n_2}\delta(t_1-t_2)\delta_{\bk,\bk_2}\delta_{\bq,\bq_2}\delta_{a,a_2}\delta_{a',a'_2}\\
-&\frac{\mathcal{A}}{e}\sum_{\bk'}
\frac{\delta}{\delta V^{a_2, a'_2}_{\textrm{eff}}(\bk_2+\bq_2,\bq_2,\underline{t}_2)}
\Big\{
\sum_{e,e'} V^{a,e,e',a'}_{\bk,\bk',\bk'-\bq,\bk+\bq} \ll e^{\dagger}_{\bk'}(\underline{t}_1) e'^{\phantom\dagger}_{\bk'-\bq}(\underline{t}_1) \gg
-\sum_{h,h'} V^{a,h,h',a'}_{\bk,\bk',\bk'-\bq,\bk+\bq} \ll h^{\dagger}_{\bk'}(\underline{t}_1) h'^{\phantom\dagger}_{\bk'-\bq}(\underline{t}_1) \gg
\Big\}\, .
\end{split}
\label{eq:DF_2}
\end{equation}
Consistent with the derivation of Hartree-type self-energy terms, we proceed by expanding the single-particle densities in terms of exciton densities according to Eq.~(\ref{eq:projection}) and interpreting the resulting four-particle expectation values as exciton Green functions. This means that Eq.~(\ref{eq:DF_2}) contains functional derivatives of the exciton Green function with respect to the effective potential, which we have identified before as the polarization function, see Eq.~(\ref{eq:deriv_K_Veff}). Thus, the dielectric function has the well-known form $\varepsilon=1-V\Pi$ with an excitonic polarization function $\Pi$.
The polarization function is calculated in RPA by inserting the delta-like vertex function (\ref{eq:vertexfct_approx}):
\begin{equation}
 \begin{split}
\Pi(3,4,\bk_2+\bq_2,\bq_2,a_2, a'_2,\underline{t}_2)\approx
\frac{e}{\mathcal{A}}\Big\{
\sum_{\bk_1,h_1} &K(3,a_2,\bk_2,h_1,\bk_1,\underline{t}_2)K(a'_2,\bk_2+\bq_2,h_1,\bk_1,\underline{t}_2,4) \\
-\sum_{\bp_1,e_1}&K(3,e_1,\bp_1,a_2,\bk_2,\underline{t}_2)K(e_1,\bp_1,a'_2,\bk_2+\bq_2,4)
\Big\}\, .
\end{split}
\label{eq:pol_RPA}
\end{equation}
Since we consider only two-particle Green functions $K$ that describe electron-hole pairs, the band index $a_2$ has to be an electron index in the first line and a hole index in the second line of Eq.~(\ref{eq:pol_RPA}), respectively. We now evaluate Eq.~(\ref{eq:DF_2}) using the RPA polarization function, introducing the exciton representation (\ref{eq:exciton_representation_inv}) and assuming that Green functions are diagonal in the exciton basis, which results in:
\begin{equation}
 \begin{split}
&\varepsilon(\bk+\bq,\bq,a,a',\underline{t}_1,\bk_2+\bq_2,\bq_2,a_2,a'_2,\underline{t}_2)\\
=&n_1\delta_{n_1,n_2}\delta(t_1-t_2)\delta_{\bk,\bk_2}\delta_{\bq,\bq_2}\delta_{a,a_2}\delta_{a',a'_2} \\
-i\hbar&\sum_{\nu_1,\nu_2,\bQ_1,\bQ_2}G(\nu_1,\bQ_1,\underline{t}_1,\nu_1,\bQ_1,\underline{t}_2)G(\nu_2,\bQ_2,\underline{t}_2,\nu_2,\bQ_2,\underline{t}^+_1)\\
\sum_{\bk'}
\Big\{
&\sum_{e,e',h'',\bk''} V^{a,e,e',a'}_{\bk,\bk',\bk'-\bq,\bk+\bq}\\
\Big(
&\sum_{h_1,\bk_1}\Phi_{\nu_1,\bQ_1}(e',\bk'-\bq,h'',\bk'')\Phi^*_{\nu_1,\bQ_1}(e''',\bk_2,h_1,\bk_1)
\Phi_{\nu_2,\bQ_2}(e'''',\bk_2+\bq_2,h_1,\bk_1)\Phi^*_{\nu_2,\bQ_2}(e,\bk',h'',\bk'')\delta_{a_2,e'''}\delta_{a'_2,e''''}\\
-&\sum_{e_1,\bp_1}\Phi_{\nu_1,\bQ_1}(e',\bk'-\bq,h'',\bk'')\Phi^*_{\nu_1,\bQ_1}(e_1,\bp_1,h''',\bk_2)
\Phi_{\nu_2,\bQ_2}(e_1,\bp_1,h'''',\bk_2+\bq_2)\Phi^*_{\nu_2,\bQ_2}(e,\bk',h'',\bk'')\delta_{a_2,h'''}\delta_{a'_2,h''''}
\Big)\\
-&\sum_{h,h',e'',\bp''} V^{a,h,h',a'}_{\bk,\bk',\bk'-\bq,\bk+\bq}\\
\Big(
&\sum_{h_1,\bk_1}\Phi_{\nu_1,\bQ_1}(e'',\bp'',h',\bk'-\bq)\Phi^*_{\nu_1,\bQ_1}(e''',\bk_2,h_1,\bk_1)
\Phi_{\nu_2,\bQ_2}(e'''',\bk_2+\bq_2,h_1,\bk_1)\Phi^*_{\nu_2,\bQ_2}(e'',\bp'',h,\bk')\delta_{a_2,e'''}\delta_{a'_2,e''''}\\
-&\sum_{e_1,\bp_1}\Phi_{\nu_1,\bQ_1}(e'',\bp'',h',\bk'-\bq)\Phi^*_{\nu_1,\bQ_1}(e_1,\bp_1,h''',\bk_2)
\Phi_{\nu_2,\bQ_2}(e_1,\bp_1,h'''',\bk_2+\bq_2)\Phi^*_{\nu_2,\bQ_2}(e'',\bp'',h,\bk')\delta_{a_2,h'''}\delta_{a'_2,h''''}
\Big)
\Big\}\, .
\end{split}
\label{eq:DF_3}
\end{equation}
Similar to the inverse dielectric matrix (\ref{eq:inv_DF_elements}), the dielectric matrix elements are identified as
\begin{equation}
 \begin{split}
\varepsilon(\bk+\bq,\bq,a,a',\underline{t}_1,\bk_2+\bq_2,\bq_2,a_2,a'_2,\underline{t}_2)&
\approx\varepsilon^{a,a'_2,a_2,a'}_{\bk,\bk_2+\bq_2,\bk_2,\bk+\bq}(\underline{t}_1,\underline{t}_2)\delta_{\bq,\bq_2}\delta_{\bk,\bk_2}\\
&=\sum_{\alpha,\beta}
\big( c_{\alpha, \bk}^{a} \big)^{*} \big( c_{\beta, \bk_2+\bq_2}^{a'_2} \big)^{*} \varepsilon^{\alpha\beta}_{-\bq}(\underline{t}_1,\underline{t}_2)c_{\beta, \bk_2}^{a_2} c_{\alpha, \bk+\bq}^{a'}\delta_{\bq,\bq_2}\delta_{\bk,\bk_2}
\,,
\end{split}
\label{eq:DF_elements}
\end{equation}
assuming again diagonality in the $\bk$-index. We can now expand the Coulomb matrix elements in terms of Wannier functions and make use of the completeness of Bloch states to derive the dielectric matrix in the Wannier representation:
\begin{equation}
 \begin{split}
 \varepsilon^{\alpha\beta}_{-\bq}(\underline{t}_1,\underline{t}_2)&=\sum_{a,a',a_2,a'_2}
  c_{\alpha, \bk}^{a}   c_{\beta, \bk+\bq}^{a'_2}  \varepsilon^{a,a'_2,a_2,a'}_{\bk,\bk+\bq,\bk,\bk+\bq}(\underline{t}_1,\underline{t}_2)
 \big( c_{\beta, \bk}^{a_2}\big)^* \big( c_{\alpha, \bk+\bq}^{a'}\big)^* \\
 &=n_1\delta_{n_1,n_2}\delta(t_1-t_2)\delta_{\alpha,\beta}
 -i\hbar \sum_{\nu_1,\nu_2,\bQ_1,\bQ_2} F(\nu_1,\bQ_1,\underline{t}_1,\nu_2,\bQ_2,\underline{t}_2)
 \sum_{\delta}V^{\alpha\delta}_{-\bq} \tilde{\Pi}^{\delta\beta}_{-\bq}(\nu_1,\nu_2,\bQ_1,\bQ_2)
\,
\end{split}
\label{eq:DF_wannier}
\end{equation}
with
\begin{equation}
 \begin{split}
F(\nu_1,\bQ_1,\underline{t}_1,\nu_2,\bQ_2,\underline{t}_2) =
G(\nu_1,\bQ_1,\underline{t}_1,\nu_1,\bQ_1,\underline{t}_2)G(\nu_2,\bQ_2,\underline{t}_2,\nu_2,\bQ_2,\underline{t}^+_1)
\,.
\end{split}
\label{eq:F}
\end{equation}
Using the momentum conservation implied by the exciton wave functions, the matrix elements of $\tilde{\Pi}$ can be explicitly calculated as
\begin{equation}
 \begin{split}
\tilde{\Pi}^{\delta\beta}_{-\bq}(\nu_1,\nu_2,\bQ_1,\bQ_2)=\delta_{\bQ_2,\bQ_1+\bq}M^{\delta}_{\nu_1,\nu_2,\bQ_1}(-\bq)\big(M^{\beta}_{\nu_1,\nu_2,\bQ_1}(-\bq)\big)^*
\,
\end{split}
\label{eq:pol_wannier_aux}
\end{equation}
with the wave function overlap
\begin{equation}
 \begin{split}
M^{\delta}_{\nu_1,\nu_2,\bQ_1}(\bq)=\sum_{\bk,e,h}
\Phi_{\nu_1,\bQ_1}(e,h,\bk)\Big\{ 
&\sum_{e'} c_{\delta, \bQ_1-\bk-\bq}^{e'}\big(c_{\delta, \bQ_1-\bk}^{e}\big)^* \Phi_{\nu_2,\bQ_1-\bq}(e',h,\bk) \\
-&\sum_{h'} c_{\delta, \bk-\bq}^{h'}\big(c_{\delta, \bk}^{h}\big)^* \Phi_{\nu_2,\bQ_1-\bq}(e,h',\bk-\bq)
\Big\}^*
\,.
\end{split}
\label{eq:pol_overlap}
\end{equation}
Finally, we calculate the retarded dielectric matrix by transforming the Keldysh dielectric matrix (\ref{eq:DF_wannier}) to physical times as discussed in the previous chapter. Assuming stationary exciton distribution functions, one can show that
\begin{equation}
 \begin{split}
i\hbar F^{\textrm{ret}}(\nu_1,\bQ_1,\nu_2,\bQ_2,\omega) =
\frac{N^{\textrm{X}}_{\nu_2,\bQ_2}-N^{\textrm{X}}_{\nu_1,\bQ_1}}{E_{\nu_2,\bQ_2}-E_{\nu_1,\bQ_1}+\hbar\omega+i\gamma}
\,.
\end{split}
\label{eq:F_omega}
\end{equation}
With this, we arrive at the final result, see Eq.~(\ref{eq:eps_ab}),
\begin{equation}
 \begin{split}
 \varepsilon^{\textrm{ret},\alpha\beta}_{\bq}(\omega)=\delta_{\alpha,\beta}-\sum_{\delta} V^{\alpha\delta}_{\bq} \Pi^{\textrm{ret},\delta\beta}_{\bq}(\omega)
\,
\end{split}
\label{eq:DF_wannier_final}
\end{equation}
with the polarization matrix
\begin{equation}
 \begin{split}
\Pi^{\textrm{ret},\delta\beta}_{\bq}(\omega)=\sum_{\nu_1,\nu_2,\bQ_1}
\frac{N^{\textrm{X}}_{\nu_2,\bQ_1-\bq}-N^{\textrm{X}}_{\nu_1,\bQ_1}}{E_{\nu_2,\bQ_1-\bq}-E_{\nu_1,\bQ_1}+\hbar\omega+i\gamma}
M^{\delta}_{\nu_1,\nu_2,\bQ_1}(\bq)\big(M^{\beta}_{\nu_1,\nu_2,\bQ_1}(\bq)\big)^*
\,.
\end{split}
\label{eq:pol_wannier_final}
\end{equation}
We have thus derived a microscopic dielectric function that describes excitonic screening in RPA corresponding to the bubble-type polarization shown in Fig.~\ref{fig:pol_bubble}. Our result is a generalization of the bound-state dielectric function derived by Röpke and Der \cite{ropke_influence_1979}, which has also been used in \cite{steinhoff_exciton_2017}. In fact, the Röpke-Der dielectric function can be considered a macroscopic limit of our result. We emphasize that a microscopic treatment of screening encoded in the matrix form of $\varepsilon^{\textrm{ret},\alpha\beta}_{\bq}(\omega)$ is essential to capture local-field effects arising due to the layered structure of the crystal unit cell. The interplay of \textit{inter}-layer and \textit{intra}-layer interactions in the TMD bilayer is thereby naturally taken into account.
For numerical calculations, we use a phenomenological damping $\gamma =$ min($10$ meV, $\hbar\omega$) in Eq.~(\ref{eq:pol_wannier_final}) to ensure the correct analytic behavior in the static limit $\omega\rightarrow 0$.

\begin{figure}
\centering
\includegraphics[width=.7\columnwidth]{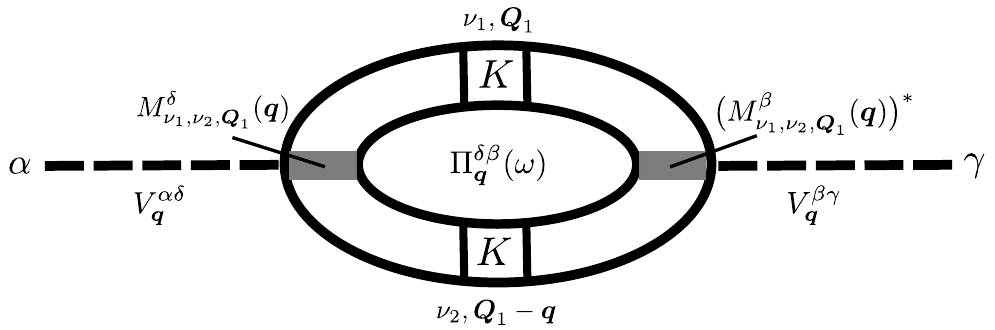}
\caption{Feynman diagram for the excitonic RPA polarization $\Pi^{\delta\beta}_{\bq}(\omega)$ inserted between two Coulomb interaction lines in Wannier representation. The interaction vertices depicted in grey correspond to the wave function overlaps defined in Eq.~(\ref{eq:pol_overlap}).}
\label{fig:pol_bubble}
\end{figure}

\subsubsection{Numerical Details for the Exciton Dyson Equation}

The exciton energy renormalizations (\ref{eq:Sigma_self_cons_main}) are evaluated self-consistently with the self-energy (\ref{eq:self-energy}) based on exciton eigenstates from the Bethe-Salpeter equation (BSE) (\ref{eq:BSE_main}). We use a Brillouin zone sampling with $48\times 48\times 1$ k-points, limiting the Brillouin zone to the regions with radius $2.3$ around the K and K' points. The two highest valence and two lowest conduction bands are considered. For every total exciton momentum $\bQ$, $48$ exciton eigenstates from the BSE are taken into account. The frequency integration in the Montroll-Ward self-energy (\ref{eq:MW_exciton}) extends from $-500$ meV to $500$ meV using a sampling with $80$ $\omega$-points. We have checked that exciton energy renormalizations are converged to within $1$ meV.

\subsubsection{Numerical Results for Spin-Triplet Excitons}

\begin{figure*}[ht]
\centering
\includegraphics[width=1.\textwidth]{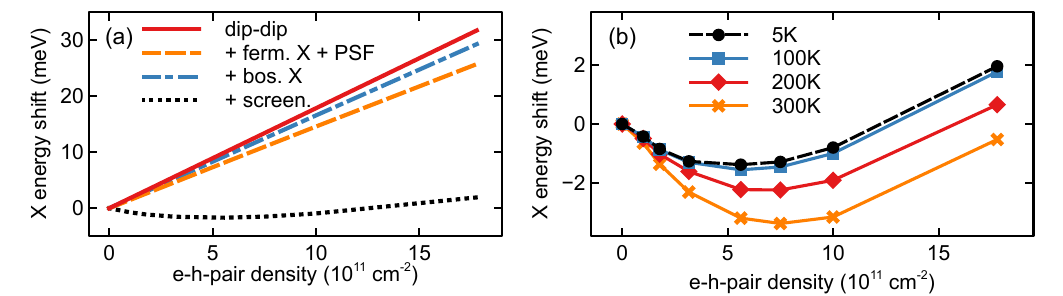}
\caption{\textbf{Exciton energy renormalizations induced by exciton-exciton interaction.} 
(a) Cumulative density-dependent renormalization of the zero-momentum ``grey'' 1s-exciton (spin-triplet) energy at a temperature $T=100$ K, subsequently adding dipole-dipole interaction (dip-dip), fermionic exchange interaction and phase-space filling (+ ferm. X + PSF), bosonic exchange interaction (+ bos. X), and screened bosonic exchange (+ screen.). The latter represents the result of the full calculation.
(b) Calculated temperature and density dependence of energy renormalization for the zero-momentum 1s-exciton triplet. 
The result for $T=5$ K has been obtained from extrapolating the high-temperature data.}
\label{fig:results_dark}
\end{figure*}

\newpage
\subsection{Experimental Details}

\subsubsection{Estimation of electron-hole pair density}
The injected electron-hole pair density was estimated using the following equation

\begin{equation}
	n_{eh} = \dfrac{P\cdot \alpha}{f_{rep} \cdot \pi\cdot r^2\cdot E_{ph}}
\end{equation}
with $P$ the laser power (ranging from 250 nW - 25 $\mu$W), $\alpha=0.11$ absorption, $f_{rep} = 80$ MHz laser repetition rate, $E_{ph}=1.63$ eV photon energy (resonant with the MoSe$_2$ A:1s state) and $r = (0.7 \pm 0.1)\,\mu$m radius of the focused laser spot (accordingly to our previous work \cite{wietek_non-linear_2023}). The absorption value was estimated from transfer matrix analysis of reflectance measurements on the heterobilayer, see Fig. \ref{fig:exp_density}. For estimating the peak density, the radius is chosen as $r = FWHM/(2 \sqrt{\ln 2})$ so that the effective density is equal to the maximum density of the spot center, $n_{eh} \times \pi r^2 = N_0$. It is assumed that the majority of the created excitons form interlayer excitons due to the ultrafast charge transfer. 
\begin{figure}[hb]
	\centering
	\includegraphics[width=.7\columnwidth]{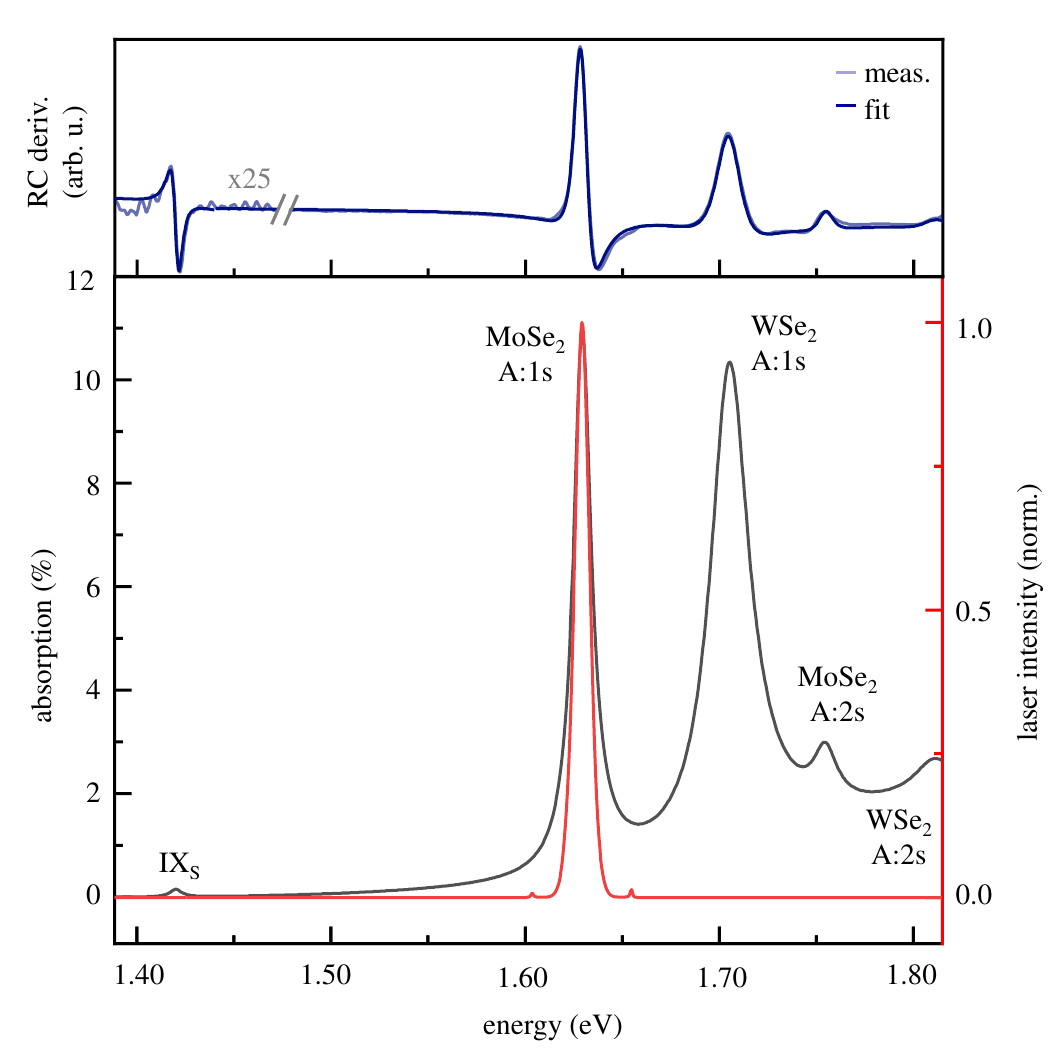}
	\caption{(Upper panel) Reflectance contrast derivative of MoSe$_2$/WSe$_2$ heterobilayer and respective fit from transfer matrix analysis. 
	The extracted parameterized dielectric function is used to obtain the effective absorption spectrum (lower panel). 
	Spectrum of the pump laser ($\lambda=761$ nm) in red, overlapping with the absorption peak of the MoSe$_2$ A:1s state. Adapted from Ref.\,\cite{wietek_non-linear_2023}.}
	\label{fig:exp_density}
\end{figure}

\subsubsection{Reproducibility across different samples}
A small blue shift was consistently observed across several different samples including CVD grown samples, gate tunable device at charge neutrality, and a sample with atomic reconstruction of R-type stacking. 
A selection of density dependent, time integrated spectra is presented in the following figures. 
All measurements were performed at a temperature of 5 K. 
Power density is given in case of cw excitation. 
A guide to the eye is inserted to visualize the small shifts in energy.
The following samples were studied:\\

\textbf{Sample A} (Fig.\ref{fig:SampleA}): Heterobilayer of H-type stacking. Missing hBN-encapsulation results in an additional inhomogeneous broadening of 10 meV. A small blue shift of 5 meV for both triplet and singlet state can be extracted. Furthermore, an indication of red shift is present.\\

\textbf{Sample B} (Fig.\ref{fig:SampleB}): Two different heterobilayers obtained by stacking-assembly of CVD-grown monolayers. The data was reported in \cite{Zhao2023} and the figure is partially adapted.\\

\textbf{Sample C} (Fig.\ref{fig:SampleC}): Encapsulated single gate heterobilayer device. Panel a)-c) show power dependent spectra with intrinsic doping (no applied gate voltage). The device was tuned into charge neutrality for panel d)-e). No charged states are detected. Additional 0D arrays from adjacent domain boundaries contribute on the low energy side (<1.39 eV). Indications of a red shift are present at low densities in panel f).\\

\textbf{Sample D} (Fig.\ref{fig:SampleD}): Encapsulated heterobilayer on CVD-grown diamond substrate. The panels refer to two different positions on the sample. Due to  sample size the positions are dislocated domains.\\

\textbf{Sample E} (Fig.\ref{fig:SampleE}): Encapsulated heterobilayer of R-type stacking on CVD-grown diamond substrate. The dominant registry is R$^X_h$ with singlet state being the lowest in energy. This stacking also shows a blue shift of only 3 meV for both charged and neutral singlet excitons.\\

\begin{figure*} [ht]
	\centering
	\includegraphics[width=.65\columnwidth]{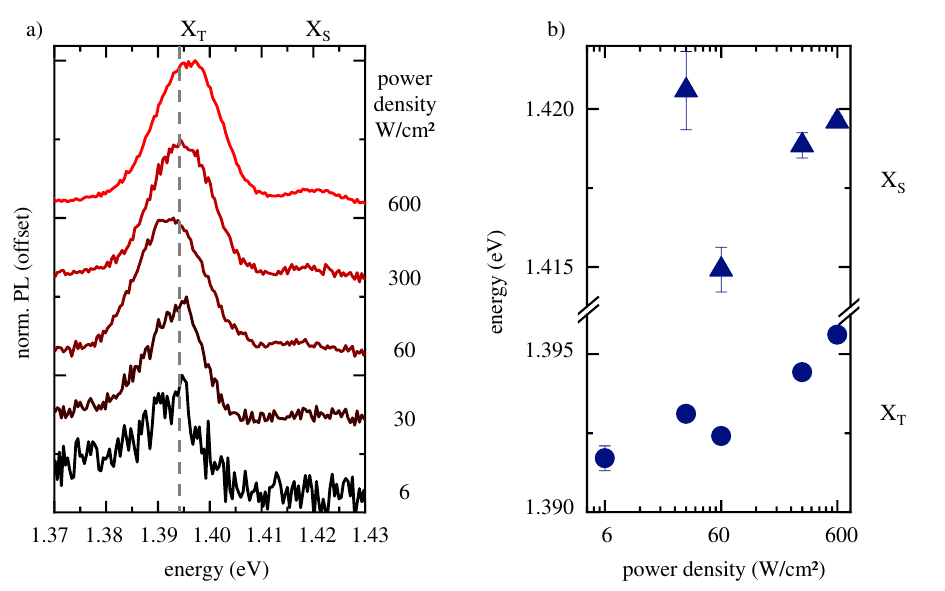}
	\caption{\textbf{Sample A:} Mechanically exfoliated, not encapsulated heterobilayer of H-type stacking. a) Time-integrated spectra spanning across two order of power density magnitude. b) Extracted blue shift of 5 meV for both triplet and singlet exciton states. 
}
	\label{fig:SampleA}
\end{figure*}
\begin{figure*} [ht]
	\centering
	\includegraphics[width=\columnwidth]{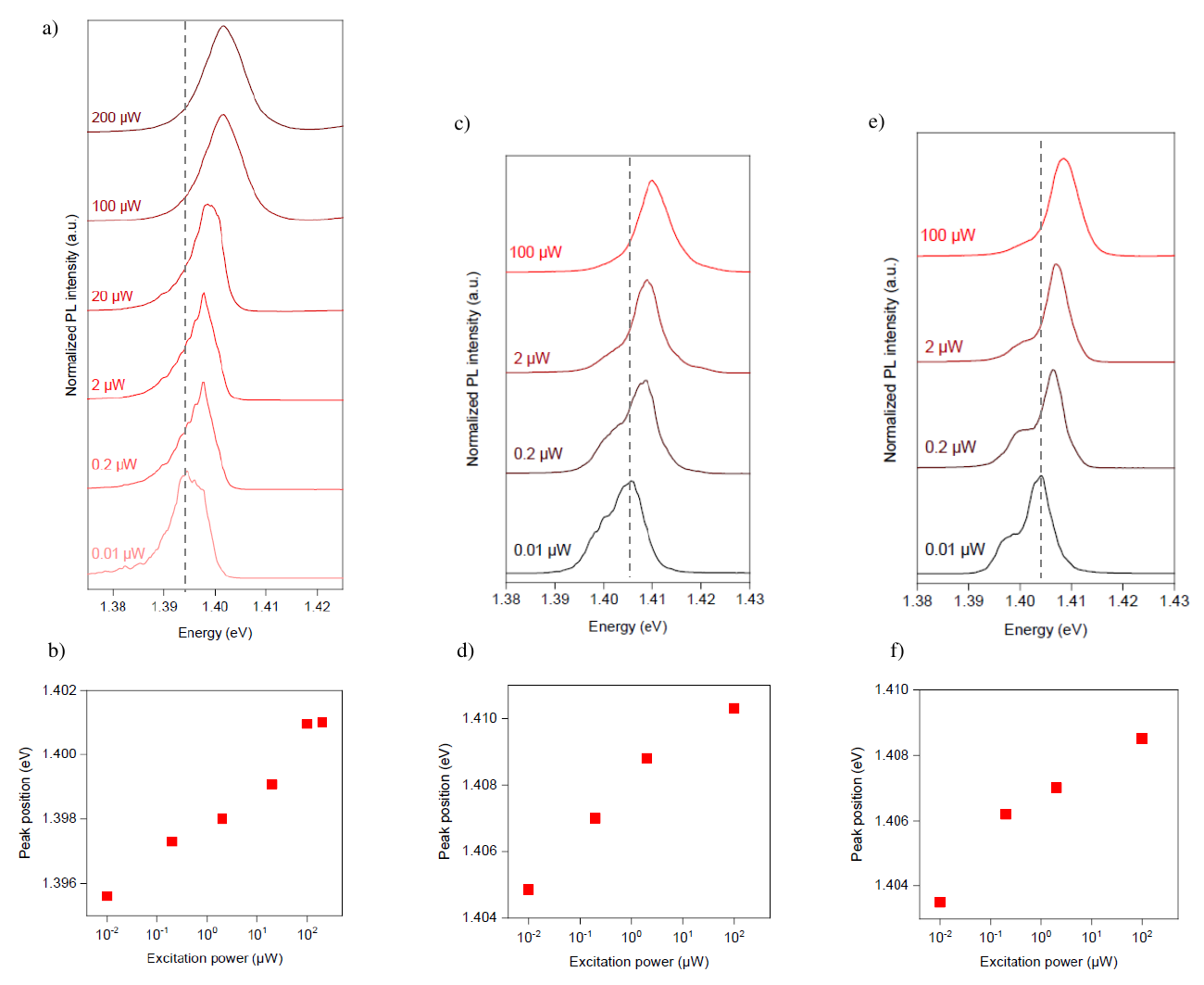}
	\caption{\textbf{Sample B:} Two samples of heterobilayers assembled from CVD-grown monolayers reproduced from \cite{Zhao2023}. Panels a) and b) correspond to the first sample, while panels c) and d) correspond to one position on the second sample and panels e) and f) correspond to another position on the second sample.}
	\label{fig:SampleB}
\end{figure*}
\newpage
~\newpage

\begin{figure*}[ht]
	\centering
	\includegraphics[width=\columnwidth]{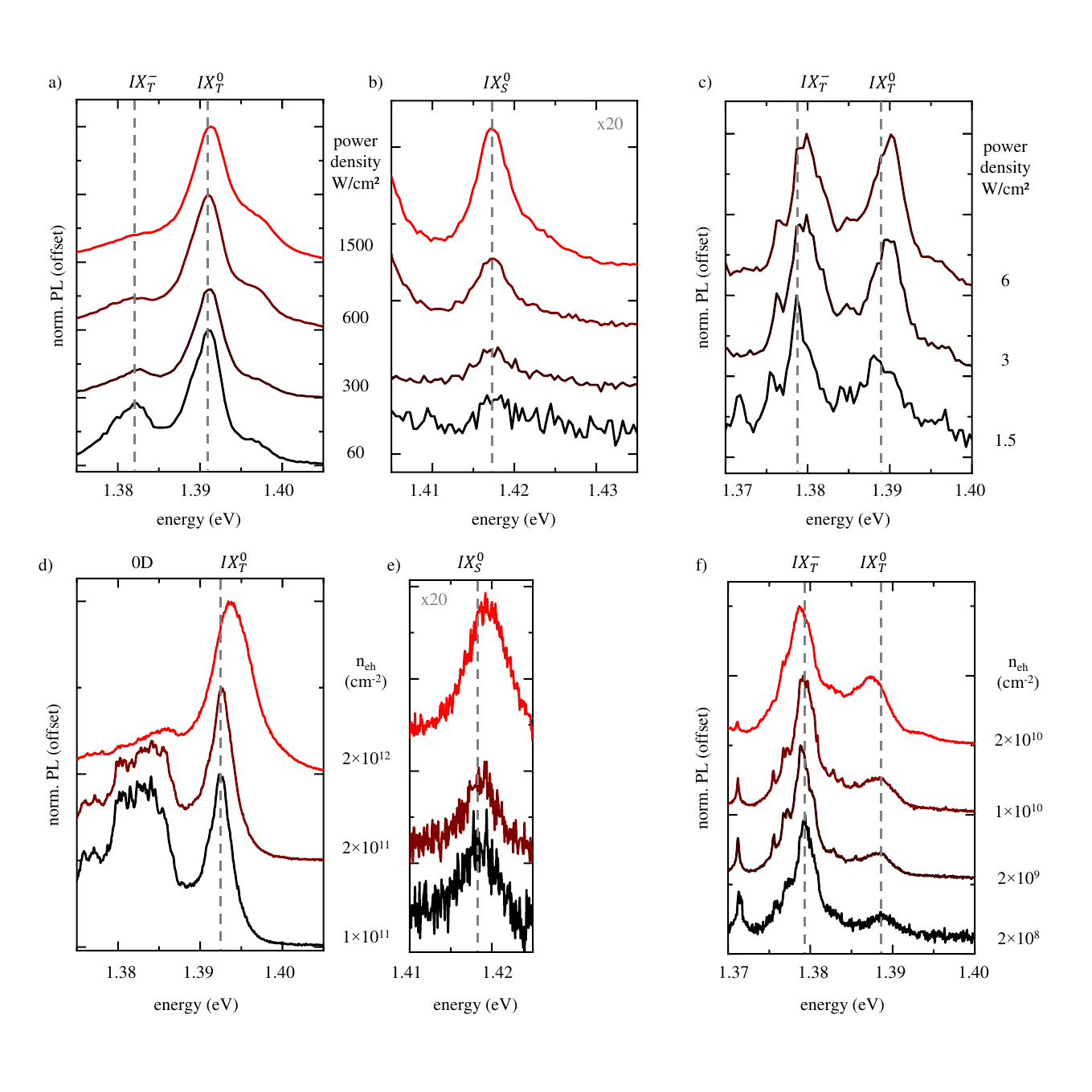}
	\caption{\textbf{Sample C:} Single gate device. a)-c) Time integrated spectra for intrinsic doping, d)-e) at charge neutrality (including 0D domain arrays), and f) low n-doping.}
	\label{fig:SampleC}
\end{figure*}
\newpage

\begin{figure} [ht]
	\centering
	\includegraphics[width=0.8\columnwidth]{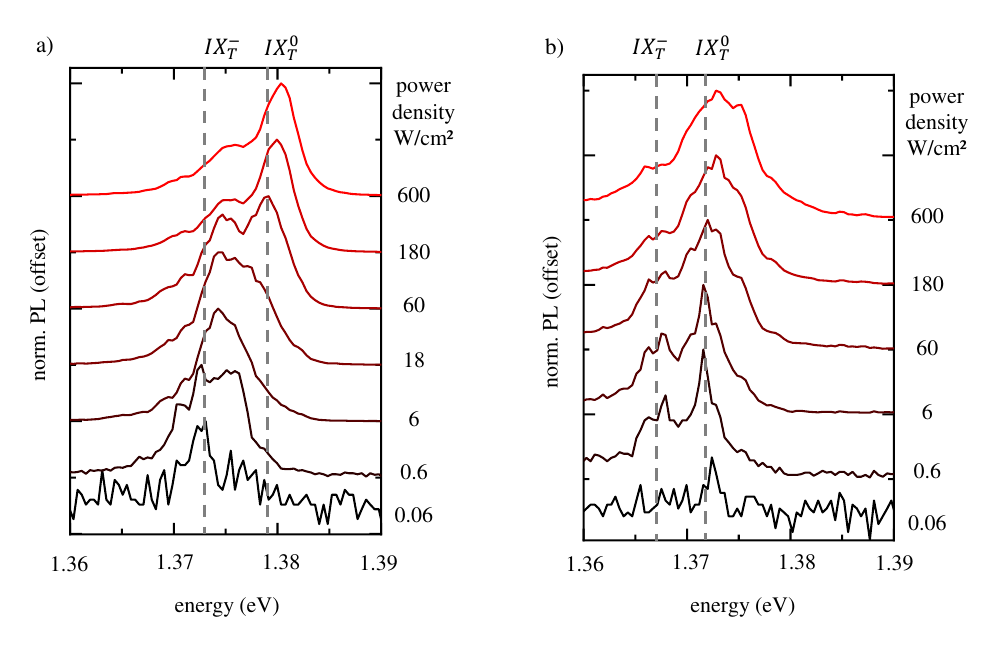}
	\caption{\textbf{Sample D:} H-type reconstructed sample on diamond substrate. a) and b) correspond to different positions on the sample.}
	\label{fig:SampleD}
\end{figure}

\begin{figure} [ht]
	\centering
	\includegraphics[width=\columnwidth]{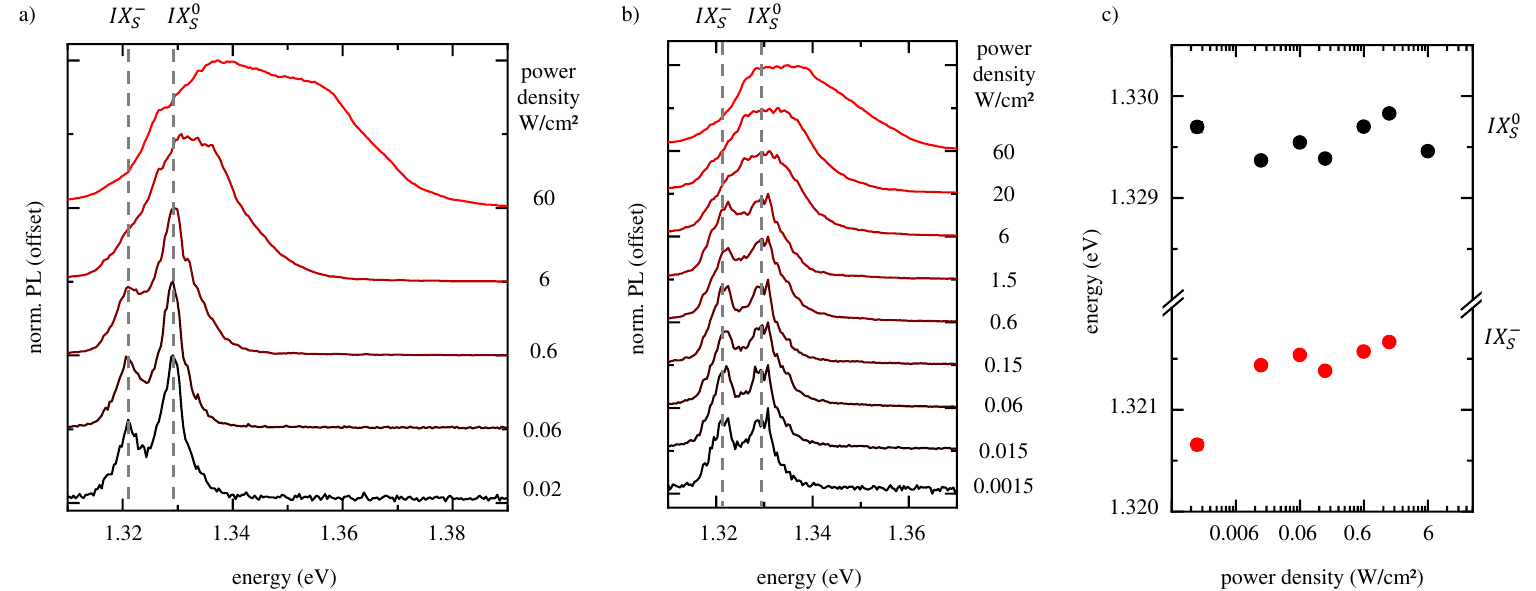}
	\caption{\textbf{Sample E:} R-type reconstruction on diamond substrate a) cw and b) pulsed excitation. No estimation of electron-hole pair density as the PL decay time exceeds the repetition rate of our laser system. c) Peak position extracted for charged and neutral singlet state.}
	\label{fig:SampleE}
\end{figure}
\newpage
\end{widetext}

\newpage
\textbf{Acknowledgements}
We acknowledge financial support from the Deutsche Forschungsgemeinschaft via SPP2244 (Project-ID: 443405595), Emmy Noether Initiative (CH 1672/1, Project-ID: 287022282), SFB 1277 (project B05, Project-ID: 314695032), the Würzburg-Dresden Cluster of Excellence on Complexity and Topology in Quantum Matter (ct.qmat) (EXC 2147, Project-ID 390858490), and the Munich Center for Quantum Science and Technology (MCQST) (EXC 2111, Project-ID 390814868), as well as resources for computational time at the HLRN (Göttingen/Berlin). 
M. F. and S. Z. acknowledge support by the Alexander von Humboldt foundation. 
K. W. and T. T. acknowledge support from the JSPS KAKENHI (Grant Numbers 20H00354, 21H05233 and 23H02052) and World Premier International Research Center Initiative (WPI), MEXT, Japan.
A. H. acknowledges funding by the
European Research Council (ERC) (Grant Agreement No. 772195).
The authors would also like to thank Mikhail M. Glazov and Sivan Refaely-Abramson for valuable discussions.



%


\end{document}